\documentclass[10pt,usletter,twocolumn]{article}
\usepackage[top=1.8cm, bottom=2cm, left=1.6cm, right=1.6cm]{geometry}

\usepackage{antpolt} %
\usepackage[QX]{fontenc}

\usepackage{amsmath,amssymb,amsfonts}
\usepackage{listings}                       %
\usepackage[noend]{algpseudocode}
\usepackage{multirow}
\usepackage{graphicx}		%
	\setkeys{Gin}{width=\textwidth, height=\textheight, keepaspectratio}
\usepackage{pifont}
\usepackage{textcomp}
\usepackage{xcolor}
\usepackage{algorithm,algorithmicx,algpseudocode,verbatim,color}
\makeatletter
\algnewcommand{\LineComment}[1]{\Statex \hskip\ALG@thistlm \(\triangleright\) #1}
\makeatother
\makeatletter
\algnewcommand{\BlankLine}[1]{\Statex \hskip\ALG@thistlm #1}
\makeatother
\usepackage{titling}
\usepackage[small,bf]{caption}
\usepackage[square,sort,comma,numbers]{natbib}

\usepackage{xcolor}
\usepackage{booktabs}
\usepackage{paralist}
\usepackage{threeparttable}

\usepackage{titlesec}
\titlespacing*{\section}{0pt}{*3}{6pt}
\titlespacing{\subsection}{0pt}{*2.5}{6pt}
\titlespacing{\subsubsection}{0pt}{*2}{2pt}

\newcommand{\descr}[1]{\medskip\noindent\textbf{#1}}

\definecolor{providercol}{HTML}{2b8a3e}
\definecolor{adversarycol}{HTML}{c92a2a}
\definecolor{linkcol}{rgb}{0,0,0.5}
\definecolor{citecol}{rgb}{0,0.5,0.3}
\definecolor{urlcol}{rgb}{0.3,0,0}

\usepackage{xspace}

\renewenvironment{thebibliography}[1]{
  \begin{oldthebibliography}{#1}
    \setlength{\itemsep}{0.0em}
    \setlength{\parskip}{0.0em}
}
{
  \end{oldthebibliography}
}

\usepackage[hang,flushmargin]{footmisc}

\renewcommand{\footnoterule}{%
  \kern -3pt
  \hrule width 1in
  \kern 2pt
}

\algrenewcomment[1]{\(\triangleright\) #1}

\usepackage[hyphens]{url}

\makeatletter
\def\url@leostyle{%
  \@ifundefined{selectfont}{\def\UrlFont{}}%
  {\def\UrlFont{}}%
}
\makeatother
\usepackage[hyphenbreaks]{breakurl}

\definecolor{darkred}{RGB}{153,0,0}
\definecolor{darkblue}{RGB}{0,0,99}
\usepackage[colorlinks=true, linkcolor=darkred, citecolor=darkred, urlcolor=darkblue]{hyperref}

\usepackage[aboveskip=2pt,belowskip=0pt]{subcaption}
\usepackage{indentfirst}

\newcommand{\sptable}{\vspace{0pt}}

\graphicspath{{images/}}
\usepackage{amsmath,amsthm,amssymb,graphicx}
\usepackage{algorithm,algorithmicx,algpseudocode,verbatim,color}
\usepackage{enumitem}
\usepackage[all]{xy}
\usepackage{flushend}
\usepackage{mathtools}
\usepackage{breqn}
\newcommand{\empeps}{\varepsilon_{emp}}
\newcommand{\approxdp}{($\varepsilon, \delta$)-DP\xspace}
\newcommand{\gdp}{$\mu$-GDP\xspace}
\newcommand{\fdp}{$f$-DP\xspace}
\newcommand{\mywidth}{0.9999}
\newcommand{\FPR}{\text{FPR}}
\newcommand{\FNR}{\text{FNR}}
\usepackage[scaled=0.95]{inconsolata}
\newcommand{\Nat}{\texttt{Target-Canary}\xspace}
\newcommand{\PI}{\texttt{Partially-Informed}\xspace}
\newcommand{\PINorm}{Partially-Informed\xspace}
\newcommand{\WC}{\texttt{Worst-Case}\xspace}

\newcommand{\SMFull}{Batched Gaussian Mechanism\xspace}
\newcommand{\SM}{\text{BGM}\xspace}
\newif\ifcomment
\commenttrue
\commentfalse
\ifcomment
	\newcommand{\edc}[1]{\textbf{\em\color{red}EDC: #1}}
	\newcommand{\sundar}[1]{\textbf{\em\color{blue}MS: #1}}
  \newcommand{\jh}[1]{\textbf{\em\color{teal}JH: #1}}
  \newcommand{\bb}[1]{\textbf{\em\color{magenta}BB: #1}}
\else
	\newcommand\edc[1]{}
	\newcommand\sundar[1]{}
  \newcommand\jh[1]{}
  \newcommand\bb[1]{}
\fi

\newtheorem{definition}{\em Definition}[section]

\begin{document}
\sloppy

\title{\bf To Shuffle or not to Shuffle: Auditing DP-SGD with Shuffling\thanks{Published at the 33rd Network and Distributed System Security Symposium (NDSS 2026) -- please cite the NDSS version.}}
\date{}

\author{Meenatchi Sundaram Muthu Selva Annamalai$^1$, Borja Balle$^2$, Jamie Hayes$^2$, Emiliano De Cristofaro$^3$\\[0.5ex]
$^1$University College London $\;\;\;^2$Google DeepMind $\;\;\;^3$UC Riverside}

\maketitle
\thispagestyle{plain}
\pagestyle{plain}

\begin{abstract}
The Differentially Private Stochastic Gradient Descent (DP-SGD) algorithm supports the training of machine learning (ML) models with formal Differential Privacy (DP) guarantees.
Traditionally, DP-SGD processes training data in batches using Poisson subsampling to select each batch at every iteration.
More recently, shuffling has become a common alternative due to its better compatibility and lower computational overhead.
However, computing tight theoretical DP guarantees under shuffling remains an open problem.
As a result, models trained with shuffling are often evaluated as if Poisson subsampling were used, which might result in incorrect privacy guarantees.\smallskip

This raises a compelling research question: can we verify whether there are gaps between the theoretical DP guarantees reported by state-of-the-art models using shuffling and their actual leakage?
To do so, we define novel DP-auditing procedures to analyze DP-SGD with shuffling and measure their ability to tightly estimate privacy leakage vis-\`a-vis batch sizes, privacy budgets, and threat models.
Overall, we demonstrate that DP models trained using this approach have considerably overestimated their privacy guarantees (by up to 4 times).
However, we also find that the gap between the theoretical Poisson DP guarantees and the actual privacy leakage from shuffling is not uniform across all parameter settings and threat models.
Finally, we study two common variations of the shuffling procedure that result in even further privacy leakage (up to 10 times).
Overall, our work highlights the risk of using shuffling instead of Poisson subsampling in the absence of rigorous analysis methods.
\end{abstract}

\section{Introduction}

To mitigate privacy risks~\cite{shokri2017membership,carlini2022membership,ye2022enhanced}, the Differentially Private Stochastic Gradient Descent (DP-SGD)~\cite{abadi2016deep} algorithm is increasingly being used to train models with Differential Privacy (DP) guarantees~\cite{dwork2006calibrating}.
More precisely, DP provably bounds the leakage from a model so that no adversary can confidently learn (up to a privacy parameter $\varepsilon$) any individual-level information about the training data.
DP-SGD is supported by many open-source libraries~\cite{yousefpour2021opacus,tfprivacy,jax-privacy2022github} and is deployed in state-of-the-art (SOTA) private models with performance increasingly approaching that of non-private models~\cite{de2022unlocking}.

\descr{Subsampling vs.~Shuffling in DP-SGD.} Since DP-SGD is computationally intensive, practitioners often attempt to optimize it.
As DP-SGD processes training data in batches, the standard approach to select batches at each step is {\em Poisson subsampling}, which requires random access to the entire dataset and can be slow for large datasets~\cite{ponomareva2023dp}.
As a result, recent work has often replaced that with {\em shuffling} the training data and deterministically iterating over fixed-size batches~\cite{de2022unlocking,tramer2021differentially,shamsabadi2024confidential}.
Additionally, modern ML pipelines (e.g., XLA compilation~\cite{xla}) are optimized for fixed batch sizes, while Poisson subsampling produces different batch sizes, which can significantly slow down private training~\cite{choquette-choo2025nearexact}.

While the privacy analysis of Poisson subsampling is well-studied and admits strong privacy amplification theorems~\cite{balle2018privacy,dong2019gaussian}, correctly estimating DP-SGD's theoretical guarantees when using shuffling remains an open problem~\cite{chua2024scalable,chua2024private,feldman2025privacy}. %
As a result, it has become common to train models using DP-SGD with shuffling while reporting DP guarantees as if Poisson subsampling were used~\cite{ponomareva2023dp,de2022unlocking,LTLH22}.
(In the rest of the paper, we use DP-SGD (Shuffle) to denote the former %
DP-SGD (Poisson) for the latter.) %
This prompts the need to analyze whether this discrepancy affects the actual privacy guarantees of state-of-the-art models using shuffling.

\begin{figure}[t]
  \centering
  \includegraphics[width=0.955\linewidth]{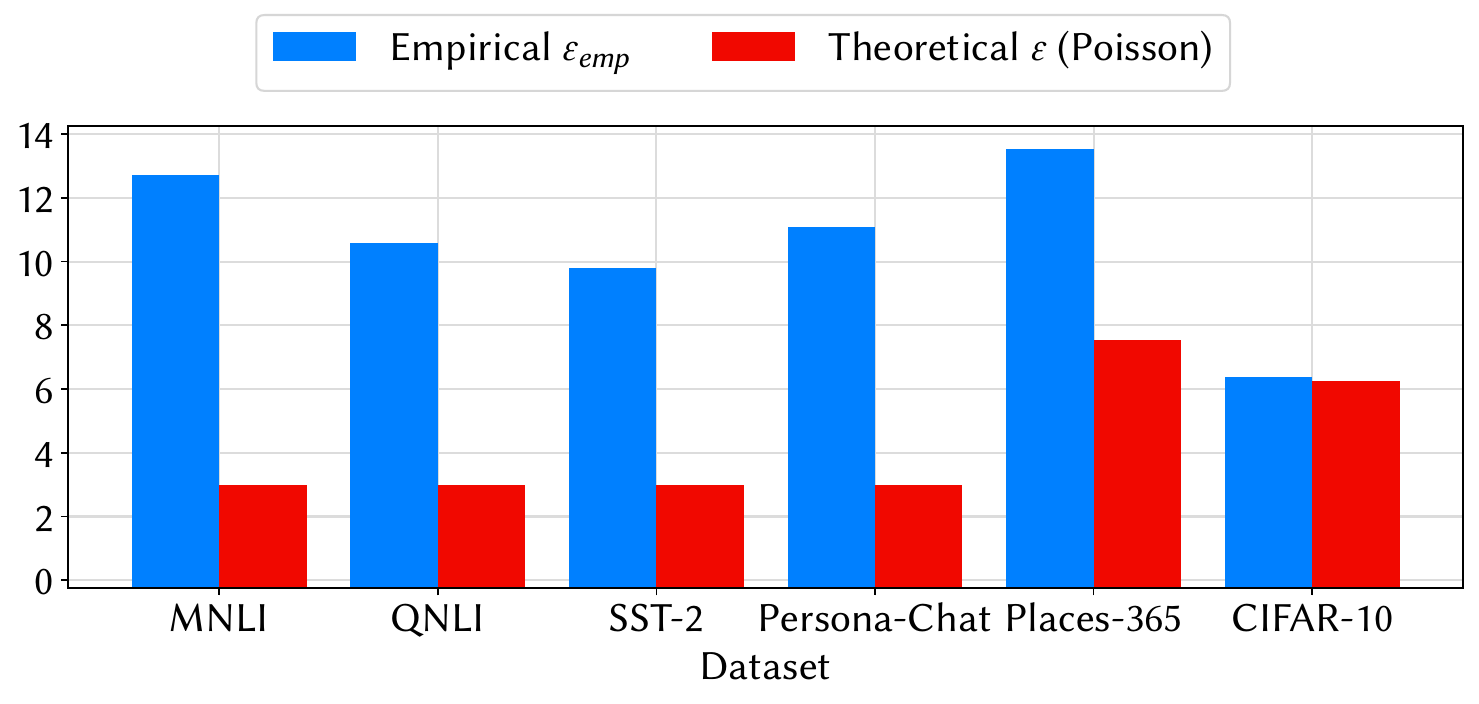}
  \caption{Largest gaps observed %
  for each dataset between the empirical privacy leakage estimate $\empeps$ and theoretical Poisson guarantees $\varepsilon$ when training by shuffling datasets using DP-SGD.} %
  \label{fig:intro_exp}
\end{figure}

\descr{DP Auditing.} We build on the concept of DP auditing~\cite{ding2018detecting}, which consists in estimating and comparing the empirical privacy leakage (denoted as $\empeps$) from DP mechanisms against their theoretical DP upper bounds ($\epsilon$)~\cite{jayaraman2019evaluating,jagielski2020auditing,nasr2021adversary,nasr2023tight,annamalai2024nearly,cebere2024tighter,galen2024oneshot}.
This involves running attacks against the mechanism -- e.g., in the case of DP-SGD, inferring whether or not a sample was used to train the model (aka membership inference~\cite{shokri2017membership}), and using the adversary's success to estimate $\empeps$. 
Finding that $\empeps > \epsilon$ indicates the presence of DP violations or bugs in the DP algorithm. %
For the auditing to be effective, it also needs not to be ``loose'' -- i.e., if $\empeps\ll\epsilon$, the audit may not be exploiting the maximum possible privacy leakage.

\descr{Research Problem.} In this paper, we present, to the best of our knowledge, the first %
methodology to audit DP-SGD (Shuffle), aiming to analyze the gap between its empirical privacy leakage and the theoretical guarantees provided by DP-SGD (Poisson).
Notably, auditing the former is significantly harder than the latter, and presenting suitable algorithms has remained a largely unaddressed research question (see Section~\ref{sec:related}).
Unlike for DP-SGD (Poisson), tight theoretical upper bounds are not currently known for DP-SGD (Shuffle)~\cite{chua2024private,chua2024scalable}, and privacy attacks have not been evaluated against it.
Therefore, it is unclear whether and how the privacy leakage can be effectively leveraged by the auditing adversary.

\descr{Roadmap.} 
We start by auditing a simplified version of DP-SGD, which we denote as \SMFull (\SM), using a novel method that builds on likelihood ratio functions.
Our \SM variant is non-adaptive and simpler than the Adaptive Batch Linear Query (ABLQ) mechanism studied in prior work~\cite{chua2024private}, enabling us to audit it tightly in a principled way (see Section~\ref{sec:dpaudit}).
Then, we extend our auditing procedure to DP-SGD (Shuffle) and evaluate the empirical privacy leakage under various adversarial models.
Our experiments highlight a substantial gap (up to $4\times$) between the empirical privacy leakage observed from SOTA models~\cite{de2022unlocking,LTLH22} and their claimed theoretical DP guarantees.
In Figure~\ref{fig:intro_exp}, we summarize the largest gaps for each dataset we experiment with -- e.g., on the MNLI dataset, the state-of-the-art differentially private BERT model from~\cite{LTLH22} reports a theoretical $\varepsilon = 3$, while our audit results in an empirical $\empeps = 12.7$ due to shuffling the dataset as opposed to using Poisson sub-sampling as in the theoretical analysis.

However, the gap is not uniform across all parameter settings and threat models.
Specifically, for large batch sizes and weak threat models, we find that the gap is much smaller, implying that the relationship between the theoretical Poisson DP guarantees and the actual privacy leakage observed from shuffling is a complex one.

Finally, we adapt our auditing procedures to audit %
two variations of the shuffling procedure first reported by Ponomareva et al.~\cite{ponomareva2023dp}, namely, partial shuffling and batch-then-shuffle. 
We find them in $2.6\%$ of public code repositories related to non-private ML training and show that these variations yield even larger privacy leakage (up to $10\times$) compared to standard shuffling.
For theoretical $\varepsilon = 0.1$, the partial shuffling and batch-then-shuffle procedures result in an empirical privacy leakage of $\empeps = 0.29$ and $\empeps = 1.00$, respectively.

\descr{Contributions.} Our work makes several contributions: %
\begin{itemize}[leftmargin=10pt]
\item We are the first to audit DP-SGD (Shuffle) and show that the empirical privacy leakage of SOTA models~\cite{de2022unlocking,LTLH22} is substantially larger than the theoretical DP guarantees, evaluating the impact of the batch size and the adversarial model on the privacy leakage observed from DP-SGD (Shuffle). %
In the process, we present novel auditing techniques %
using likelihood ratio functions.\smallskip
\item We investigate and identify gaps in the privacy leakage from two common variants of the shuffling procedure, namely, partial shuffling and batch-then-shuffle. \smallskip
\item Although we focus on DP-SGD (Shuffle), our auditing framework could be applied to any sampling technique and mechanism, e.g., to identify variations within and estimate privacy leakage for implementations of shuffling itself.\smallskip
\item Our work attests to the impact of shuffling (and its variants) on privacy leakage, calling into question the guarantees claimed by some SOTA models.
This has important implications %
both in terms of privacy (i.e., privacy claims might be overly optimistic) and utility (i.e., hyperparameters tuned to DP-SGD (Shuffle) may not be optimal for DP-SGD (Poisson) and vice versa).
\end{itemize}

\noindent {\em NB:} we have made the source code of our auditing procedure publicly available to support reproducibility and encourage further work in this space.\footnote{See \url{https://github.com/spalabucr/audit-shuffle}.}

\section{Background}

\subsection{Differential Privacy (DP)}

\begin{definition}[Differential Privacy (DP)~\cite{dwork2006calibrating}]
  \label{def:dp}
  A randomized mechanism $\mathcal{M} : \mathcal{D} \rightarrow \mathcal{R}$ is $(\varepsilon, \delta)$-differentially private if for any two adjacent datasets $D, D' \in \mathcal{D}$ and $S \subseteq \mathcal{R}$:

{\small\begin{equation*}
    \Pr[\mathcal{M}(D) \in S]  \leq e^\varepsilon \Pr[\mathcal{M}(D') \in S] + \delta
  \end{equation*}}
\end{definition}
Two common notions of adjacent datasets are {\em add/remove} and {\em edit}. %
The former corresponds to inserting/deleting a single record from the dataset (hence $|D| = |D'| \pm 1$); the latter to replacing a single record with another ($|D| = |D'|$).
Note that the difference in the size of the adjacent datasets %
limits the applicability of the add/remove adjacency in some settings (e.g., sampling w/o replacement), while guarantees under edit adjacency typically require roughly twice the amount of noise~\cite{ponomareva2023dp} since replacing a record is equivalent to first deleting then adding a record under the add/remove adjacency.

\descr{Zero-out Adjacency.}
To bridge the gap between add/remove and edit adjacency, the ``zero-out'' adjacency~\cite{kairouz2021practical} is increasingly used to simplify theoretical privacy analyses~\cite{chua2024private,choquette2024amplified,ponomareva2023dp}.

\begin{definition}[Zero-out adjacency~\cite{kairouz2021practical}]
  \label{def:zero_out}
  Let $\mathcal{X}$ be a data domain s.t.~special element $\bot \notin \mathcal{X}$ and $\mathcal{X}_\bot = \mathcal{X} \cup \{\bot\}$.
  Datasets $D \in \mathcal{X}^n$ and $D' \in \mathcal{X}_\bot^n$ are zero-out adjacent if exactly one record in $D$ is replaced with $\bot$ in $D'$.
\end{definition}

The special $\bot$ record %
is usually 0 for numerical data.
This allows the guarantees under zero-out adjacency to be semantically equivalent to those under add/remove (i.e., no additional noise required) while ensuring that the sizes of the adjacent datasets are equal.

\descr{\fdp and trade-off functions.}
Besides \approxdp, there are other formalizations of DP; %
e.g., \fdp captures the difficulty for any adversary to distinguish between the outputs of a mechanism $\mathcal{M}$ on adjacent datasets $D$ and $D'$ using trade-off functions.

\begin{definition}[Trade-off function~\cite{dong2019gaussian}]
  \label{def:tradeoff_fn}
  For any two probability distributions $P$ and $Q$ on the same space, the trade-off function $T(P, Q) : [0, 1] \rightarrow [0, 1]$ is defined as:

  {\small\begin{equation*}
    T(P, Q) (\alpha)= \inf \{ \beta_\phi : \alpha_\phi \leq \alpha \}
  \end{equation*}}
  where the infimum is taken over all (measurable) rejection rules $\phi$ and $\alpha_\phi$ and $\beta_\phi$ are the type I and II errors corresponding to this rejection rule, respectively.
\end{definition}
In theory, the trade-off function characterizes the false positive/negative errors achievable by any adversary aiming to distinguish between $P$ and $Q$.
A mechanism $\mathcal{M}$ is said to satisfy \fdp if for all adjacent datasets $D$, $D'$:
 $ T(\mathcal{M}(D), \mathcal{M}(D')) \geq f $.

One special case of \fdp is \gdp, when the underlying distributions are Gaussian.
\begin{definition}[\gdp~\cite{dong2019gaussian}]
  \label{def:gdp}
  A mechanism $\mathcal{M}$ satisfies \gdp if for all adjacent datasets $D$, $D'$:

{\small  \begin{equation*}
    T(\mathcal{M}(D), \mathcal{M}(D')) \geq \Phi(\Phi^{-1}(1 - \alpha) - \mu),
  \end{equation*}}
  $\Phi$ being the standard normal cumulative distribution function.
\end{definition}

\noindent \fdp is useful, e.g., in the context of auditing (introduced in Section~\ref{sec:dpaudit}), as it is equivalent to \approxdp for specific trade-off functions and can be used to efficiently compare the empirical power of adversaries with theoretical guarantees, as we do later in the paper.

\descr{Privacy Loss Distribution (PLD).}
The PLD formalism~\cite{koskela2020computing} is useful to derive tight theoretical guarantees for DP mechanisms.
In recent work~\cite{nasr2023tight}, PLD has also been used to tightly audit DP-SGD (Poisson) by using the given PLD to estimate the corresponding trade-off function.
However, while the PLD for DP-SGD (Poisson) is known, deriving the PLD for DP-SGD (Shuffle) is still an active area of research~\cite{chua2024private,chua2024scalable}.

\subsection{DP-SGD}

Differentially Private Stochastic Gradient Descent (DP-SGD)~\cite{abadi2016deep} is a popular algorithm for training machine learning (ML) models with formal privacy guarantees.
DP-SGD takes in input a dataset $D$ along with several hyperparameters and proceeds iteratively, processing the dataset in batches and computing the gradients of samples one batch at a time.
There are several strategies for sampling a batch from a dataset, such as Poisson subsampling and sampling without replacement.
In general, DP-SGD can be defined with an abstract batch sampler $\mathcal{B}$ that takes as input the dataset $D$ and the (expected) batch size $B$, and outputs batches sampled from the dataset.
We report its pseudo-code in Algorithm~\ref{alg:dpsgd} in Appendix~\ref{app:dpsgd}.
\descr{Poisson subsampling.} The first implementations of DP-SGD used batches sampled through Poisson subsampling.
With this approach, the batch sampler independently samples each record $(x, y) \in D$ with probability $q = B / |D|$ in each batch, where $B$ is the batch size.
One major advantage of using Poisson subsampling is substantially reducing the amount of required noise via strong privacy amplification theorems~\cite{abadi2016deep,dong2019gaussian}.
However, in practice, Poisson subsampling can be quite inefficient. %
For instance, the dataset cannot be fully loaded in memory when it is too large; thus, one needs to load a random batch in and out from the disk at every step, which is costly~\cite{ponomareva2023dp}.
Furthermore, efficient modern ML pipelines (e.g., XLA compilation) require fixed-size batches to fully leverage GPU parallelization%
~\cite{choquette-choo2025nearexact}.

\descr{Shuffling.} In practice, state-of-the-art DP-SGD implementations often rely on more computationally efficient sampling schemes, such as shuffling, while reporting DP guarantees as though Poisson subsampling was used~\cite{de2022unlocking}.
In shuffling, the batch sampler randomly permutes the records first, then partitions the dataset into fixed-size batches.
Since shuffling is already the standard in non-private training, this also simplifies implementations of DP-SGD by layering DP on top of existing non-private pipelines instead of redesigning and optimizing them from scratch.
However, since the privacy guarantees provided by shuffling are not yet fully understood, this can create a discrepancy between the theoretical guarantees reported by state-of-the-art models and the actual privacy leakage, which we study in this work.

\descr{Zero-out Adjacency.}
While shuffling is more compatible with edit adjacency due to the fixed batch sizes, the Poisson scheme is more suitable for add/remove adjancency.
Nevertheless, it is possible to derive guarantees for Poisson under the edit adjacency, although this is known to be very cumbersome~\cite{balle2018privacy}.
Thus, comparing privacy guarantees between the different subsampling schemes is inherently complicated.
Overall, zero-out adjacency is increasingly used to analyze the privacy of DP-SGD in practice~\cite{ponomareva2023dp,choquette2024amplified,kairouz2021practical}.

Specifically, the $\bot$ record is defined so that the gradient is always zero, i.e., $\forall \theta\; \nabla \ell (\bot; \theta) = \mathbf{0}$~\cite{kairouz2021practical}.
By doing so, we ensure that DP-SGD (Poisson) under zero-out adjacency is semantically equivalent to DP-SGD (Poisson) under add/remove, which enables us to compare the privacy guarantees of DP-SGD (Poisson) with that of DP-SGD (Shuffle) under a common adjacency notion. %

\section{DP Auditing}\label{sec:dpaudit}

Privacy auditing broadly denotes the process of empirically estimating the privacy leakage from an algorithm.
In DP, this involves running experiments to estimate the empirical privacy guarantees ($\empeps$) and compare them to the theoretical guarantees ($\varepsilon$).
For simplicity, we assume the empirical guarantees are derived for the same $\delta$ as the theoretical \approxdp; thus, we are technically comparing ($\empeps, \delta$) to ($\varepsilon, \delta$), but leave out $\delta$ to ease presentation.
If one finds that $\empeps>\varepsilon$, the mechanism leaks more privacy than expected.
As mentioned earlier, if $\empeps\ll\varepsilon$, the audit is not {\em tight}, i.e., the theoretical bounds are overly conservative or there may be substantial room for improvement to the privacy estimation procedure.

Implementing DP algorithms correctly is challenging~\cite{opacusbug,prngreuse} %
and bugs in DP-SGD implementations have resulted in significantly degraded protections~\cite{prngreuse,cebere2024tighter} and realistic privacy attacks~\cite{debenedetti2024privacy}.
DP auditing can be used to detect these bugs~\cite{tramer2022debugging,nasr2023tight,annamalai2024what} and/or evaluate the optimality of the attacks~\cite{annamalai2024what,nasr2021adversary}.
Moreover, empirical guarantees are also useful for estimating the privacy loss %
in settings where tight theoretical guarantees are not currently known, e.g., when the adversary does not have access to intermediate updates~\cite{galen2024oneshot,annamalai2024nearly}.

In the rest of this section, we introduce a general DP auditing procedure using an adversary and a distinguishing game. %
The attack's success %
can then be converted into the lower-bound empirical privacy leakage estimate $\empeps$. %

\subsection{The Distinguishing Game}

\begin{figure}[t]
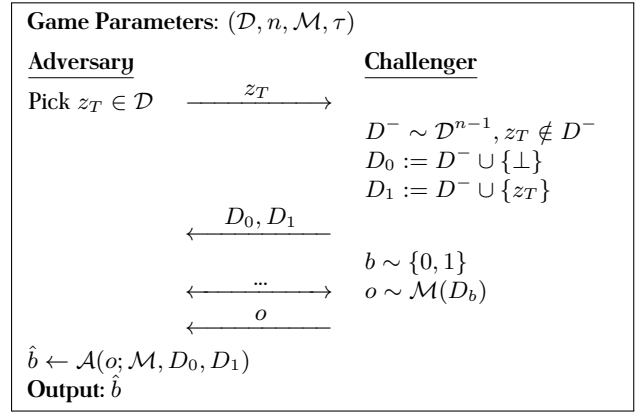

    \centering
    \fbox{\small
    \centering
    \begin{minipage}{0.93\columnwidth}
      \hspace{0.1cm}{\bf Game Parameters}: $(\mathcal{D}, n, \mathcal{M}, \tau)$\\[1ex]
      \begin{tabular}{lll}
        \hspace*{-0.1cm}\textbf{\underline{Adversary}} & & \textbf{\underline{Challenger}}\\[0.5ex]
        \hspace*{-0.1cm}Pick $z_T \in \mathcal{D}$~~~ & $\xrightarrow{\makebox[50pt]{$z_T$}}$~~ & \\
        & & $D^- \sim \mathcal{D}^{n - 1}, z_T \notin D^-$ \\
        & & $D_0 := D^- \cup \{\bot\}$ \\
        & & $D_1 := D^- \cup \{z_T\}$ \\
        & $\xleftarrow{\makebox[50pt]{$D_0, D_1$}}$ & \\
        & & $b \sim \{0, 1\}$ \\
        & $\xleftrightarrow{\makebox[50pt]{...}}$ & $o \sim \mathcal{M}(D_b)$ \\
        & $\xleftarrow{\makebox[50pt]{$o$}}$ & \\[1ex]
      \end{tabular}
      \hspace*{0.1cm}$\hat{b} \leftarrow \mathcal{A}(o; \mathcal{M}, D_0, D_1)$\\
      \hspace*{0.1cm}$\text{\bf Output:~} \hat{b} $
    \end{minipage}
  }
  \vspace{0.1cm}
  \caption{Distinguishability Game between an Adversary and a Challenger for zero-out DP given data distribution $\mathcal{D}$ of size $n$, the mechanism $\mathcal{M}$, and a decision threshold $\tau$.
  }
  \label{fig:dg}
\end{figure}

In Figure~\ref{fig:dg}, we present the standard distinguishing game~\cite{nasr2023tight}, involving an Adversary and a Challenger, adapted to audit the zero-out DP guarantees of a mechanism $\mathcal{M}$.
The Challenger runs the mechanism on a randomly chosen dataset, and the Adversary aims to determine the dataset used by the Challenger based on the mechanism's output.

In each game, the Adversary picks a single target record $z_T$ from the data domain and sends it to the Challenger.\footnote{We are abusing notation by define $\mathcal{D}$ as both the data \emph{distribution} and \emph{domain}, although it is clear from the context.}
The latter constructs adjacent datasets $D_0$ and $D_1$ by appending the zero-out record $\bot$ and the target record $z_T$, respectively, to a randomly sampled dataset with $n {-} 1$ records as done in prior work~\cite{nasr2023tight,steinke2024privacy,galen2024oneshot,pillutla2024unleashing}.
Then, the Adversary is given access to the randomly sampled adjacent datasets $D_0$ and $D_1$.
Next, the Challenger %
runs $\mathcal{M}$ on dataset $D_b$ for a random $b\in\{0,1\}$, and the Adversary wins if they correctly guess $\hat{b} = b$.

Note that $\mathcal{M}$ can be a complex algorithm with many implementation details that the theoretical privacy analysis does not rely on; thus, depending on the threat model considered and the adversarial capabilities, the adversary may be given access to specific internals of $\mathcal{M}$ as discussed in Section~\ref{sec:threat_models}, along with $M$'s output.
In theory, this can take any form, e.g., scalar $o \in \mathbb{R}$, vector $o \in \mathbb{R}^d$, or even have arbitrary domain $o \in \mathcal{Y}$.
However, it can be difficult to consider and design decision functions around values in arbitrary domains.
Hence, a common strategy~\cite{nasr2021adversary,jagielski2020auditing} is for the Adversary to define a distinguishing function $\mathcal{A}$ that assigns a scalar ``score'' to the output representing the Adversary's confidence that the output is drawn from processing $D_1$.
This score is then thresholded to produce a guess $\hat{b} = 1$ if the Adversary determines that $o \sim \mathcal{M}(D_1)$ and $0$ otherwise.

\subsection{Estimating $\empeps$}
\label{sec:estimate_eps}
We estimate the empirical privacy leakage $\empeps$ by running the distinguishing game multiple times, where only a single record is inserted in each run.
We compute the false positive rate (FPR) $\alpha$ and the false negative rate (FNR) $\beta$ across multiple games and derive %
(statistically valid) upper bounds $\overline{\alpha}$ and $\overline{\beta}$ using Clopper-Pearson confidence intervals (CIs) to quantify the confidence in our privacy loss estimation~\cite{nasr2023tight,jagielski2020auditing}.
We then calculate the empirical lower-bound privacy guarantee $\empeps$ from $\overline{\alpha}$ and $\overline{\beta}$ using the \approxdp definition as follows.

While recent DP-SGD audits have presented ``auditing in one run'' methods~\cite{steinke2024privacy,mahloujifar2024auditing,xiang2025privacy}, these do not yield sufficiently tight DP-SGD audits (even under powerful threat models), and thus we choose to audit using multiple runs. 
Moreover, prior work~\cite{nasr2023tight,mahloujifar2024auditing,xiang2025privacy} has proposed alternative auditing techniques for DP-SGD (Poisson) using \fdp or PLD; however, since tighter \fdp guarantees for DP-SGD (Shuffle) are currently unknown~\cite{chua2024private}, we stick to the \approxdp definition.

\descr{Auditing using \approxdp.}
For any given \approxdp mechanism, the possible false positive rates ($\alpha$) and false negative rates ($\beta$) attainable by any adversary are known to be the following privacy region~\cite{kairouz2015composition}:

{\small
\begin{equation}
\begin{split}
  \mathcal{R}(\varepsilon, \delta) = \{ (\alpha, \beta) | & \alpha + e^\varepsilon \beta \geq 1 - \delta \land e^\varepsilon \alpha + \beta \geq 1 - \delta~\land \\
  & \alpha + e^\varepsilon \beta \leq e^\varepsilon + \delta \land e^\varepsilon \alpha + \beta \leq e^\varepsilon + \delta \}
\end{split}
\end{equation}}

\noindent Therefore, given upper bounds $\overline{\alpha}$ and $\overline{\beta}$, the empirical lower bound %
 can be calculated as:

{\small
\begin{equation}\label{eqn:empeps}
  \empeps = \max \left\{\ln \left( \frac{1 - \overline{\alpha} - \delta}{\overline{\beta}} \right), \ln \left( \frac{1 - \overline{\beta} - \delta}{\overline{\alpha}} \right), 0 \right\}
\end{equation}\vspace{-0.1cm}}

While Bayesian intervals%
~\cite{zanella2023bayesian} can improve the privacy estimation's tightness, they are also more computationally expensive to derive and may not always be statistically sound~\cite{nasr2023tight}.
Thus, we stick to Clopper-Pearson CIs.
(In Section~\ref{sec:compare_tradeoff}, we will also evaluate the impact of using CIs on the empirical privacy leakage estimation).

For simplicity, we follow standard practice~\cite{jagielski2020auditing,zanella2023bayesian,nasr2023tight,steinke2024privacy} and fix the the value of $\delta = 10^{-5}$ throughout all experiments and only estimate the $\empeps$ guarantee at this $\delta$ value.

\subsection{The Distinguishing Function}\label{sec:distinguishing}

As mentioned, DP auditing involves an adversary distinguishing between observations from $\mathcal{M}(D)$ and $\mathcal{M}(D')$.
For a mechanism $\mathcal{M} : \mathcal{X} \rightarrow \mathcal{Y}$, this can be any function of the form $\mathcal{A}: \mathcal{Y} \rightarrow \{0, 1\}$ where $\mathcal{A}(o) = 1$ (or 0) represents that the adversary predicts that the observation $o$ is drawn from $\mathcal{M}(D)$ (or $\mathcal{M}(D')$).

\descr{Scalar Score.} In theory, $\mathcal{A}$ is equivalent to the rejection rule from the definition of $f$-DP, which represents the adversary performing a hypothesis test on whether to reject that $o \sim \mathcal{M}(D')$.
However, in previous work~\cite{jayaraman2019evaluating,jagielski2020auditing,nasr2021adversary,nasr2023tight}, the adversary assigns a scalar ``score'' to each output, which is then thresholded to form the distinguishing function---e.g., Jagielski et al.~\cite{jagielski2020auditing} and Nasr et al.~\cite{nasr2023tight} use, respectively, the loss function and dot product. %
This is because the raw outputs of %
DP-SGD tend to be high-dimensional vectors where distances can be difficult to interpret, thus making distinguishing functions hard to design.

On the other hand, using a scalar score, the adversary can first represent the \textit{confidence} that the observation was drawn from $\mathcal{M}(D)$ instead of $\mathcal{M}(D')$.
Then, the score can be thresholded to produce a prediction easily, i.e., all observations with score $\geq \tau$ are labeled as drawn from $\mathcal{M}(D)$ and from $\mathcal{M}(D')$ otherwise.
Furthermore, this threshold $\tau$ can also be adjusted to produce not only a single FPR/FNR pair but an FPR-FNR curve, i.e., an empirical \emph{trade-off curve}, which can also be compared with the claimed theoretical trade-off function when auditing.

\descr{Choosing the threshold.} When auditing using scalar scores, choosing the threshold $\tau$ from an independent set of observations is the easiest method to ensure a technically valid lower bound for the empirical privacy leakage estimate $\empeps$ derived this way.
However, it is standard to report the maximum $\empeps$ for the optimal threshold~\cite{annamalai2024nearly,cebere2024tighter,jagielski2020auditing,nasr2021adversary,nasr2023tight,pillutla2024unleashing,zanella2023bayesian} and we do so too, making the process more efficient by first sorting the scalar scores. %
Here, we substantially improve the efficiency of this procedure by first sorting the scores in increasing order, as reported in Algorithm~\ref{alg:estimate_eps}.
This enables us to find the optimal threshold from, potentially, billions of observations without incurring a prohibitive computational cost.
For simplicity, we abstract the details of choosing an appropriate threshold and designing a distinguishing function.
We refer to the empirical estimation procedure at significance level $\alpha$ and privacy parameter $\delta$ from scores $\mathcal{S}$ and $\mathcal{S}'$ derived from $\mathcal{M}(D)$ and $\mathcal{M}(D')$, respectively as $\text{EstimateEps}(\mathcal{S}, \mathcal{S}', \alpha, \delta)$.

\begin{algorithm}[t]
    \small
    \caption{Estimating $\empeps$ from scores}\label{alg:estimate_eps}
    \begin{algorithmic}[1]
    \Require Scores from $\mathcal{M}(D)$, $\mathcal{S}$. Scores from $\mathcal{M}(D')$, $\mathcal{S}'$. Significance level, $\alpha$. Privacy parameter, $\delta$.

    \LineComment Assume $|\mathcal{S}| = |\mathcal{S}'|$.
    \State $R \leftarrow |\mathcal{S}|$

    \LineComment Initial $\FPR$ and $\FNR$ for classifying all scores as $s \sim \mathcal{M}(D)$.
    \State $\FPR \leftarrow R$
    \State $\FNR \leftarrow 0$

    \For{$\tau \in \text{sort\_increasing}(\mathcal{S} \cup \mathcal{S}')$}
      \LineComment Calculate $\FPR$ and $\FNR$ for classifying score as $s \sim \mathcal{M}(D)$ if $s > \tau$.
      \If{$\tau \in S$}
        \State $\FNR \leftarrow \FNR + 1$
      \Else
        \State $\FPR \leftarrow \FPR - 1$
      \EndIf
      \LineComment Calculate confidence intervals for $\FPR$ and $\FNR$.
      \State $\overline{\FPR} \leftarrow \text{Clopper-Pearson}(FPR, R, \alpha)$
      \State $\overline{\FNR} \leftarrow \text{Clopper-Pearson}(FNR, R, \alpha)$

      \LineComment Estimate $\empeps$ from Equation~\ref{eqn:empeps}
      \State $\empeps[\tau] = \max \left\{\ln \left( \frac{1 - \overline{\FPR} - \delta}{\overline{\FNR}} \right), \ln \left( \frac{1 - \overline{\FNR} - \delta}{\overline{\FPR}} \right), 0 \right\}$
    \EndFor

    \State \Return $\max_\tau \empeps[\tau]$
    \end{algorithmic}
\end{algorithm}

\section{Auditing the \SMFull}
\label{sec:audit_sm}

Before auditing DP-SGD (Shuffle), we first turn to the \SMFull (\SM) (see Algorithm~\ref{alg:simple_mech} in Appendix~\ref{app:dpsgd}), a heavily {\em simplified} version of DP-SGD adapted from~\cite{chua2024private}.
We do so to develop principled, (close to) tight auditing techniques for shuffling (under an idealized setting).

Like DP-SGD, \SM also proceeds iteratively, sampling batches according to some batch sampler $\mathcal{B}$.
However, instead of calculating gradients, the inputs are simply aggregated together with noise at each batch, and the noisy aggregated values are released for each batch across all epochs.
Moreover, unlike DP-SGD, it is \textit{non-adaptive}, i.e., the outputs of previous iterations do not affect the current one.
While \SM can be instantiated with any batch sampler, we use the {\em shuffle} batch sampler, which is commonly used in ML training.
We bound the inputs, i.e. $\forall i\; x_i \in [-1, +1]$, so that the mechanism satisfies $(\varepsilon, \delta)$-DP.
When the number of epochs is 1, Chua et al.~\cite{chua2024private} use the adjacent datasets $D = (+1, -1, ..., -1)$ and $D' = (\bot, -1, ..., -1)$, setting $\bot = 0$ in their lower bound privacy analysis.
Specifically, they conjecture, but do not prove, that these are the worst-case adjacent datasets for the shuffle batch sampler.
Note that this is different from the proven worst-case datasets $D = (+1, 0, ..., 0)$ and $D' = (\bot, 0, ..., 0)$ for the Poisson batch sampler~\cite{zhu2022optimal}.
Specifically, in the shuffle setting, the differing sample in the conjectured worst-case datasets is equal in magnitude but opposite in direction to the other samples, unlike in the Poisson setting.
Finally, in a single epoch and for batch size $B$, the outputs of the mechanism on the adjacent datasets are $(-B, ..., -B + 2, ..., -B) + \mathcal{N}(0, \sigma^2\mathbb{I})$ and $(-B, ..., -B + 1, ..., -B) + \mathcal{N}(0, \sigma^2\mathbb{I})$ for $D$ and $D'$, respectively.
Since the inputs are shuffled, the ``$-B + 2$'' and ``$-B + 1$'' values can appear in any batch.

We stress that \SM is a ``hypothetical'' algorithm, i.e., it is not used/associated with common data distributions or training datasets.
Therefore, when auditing it using the Distinguishability Game, we take the data distribution to be the ``pathological'' distribution, which always outputs `-1.'
Then, the adversary always chooses `+1' as the target record $z_T$.

\subsection{From Single to Multiple Epochs}

\descr{Single Epoch.}
To audit a single epoch, %
an adversary has to distinguish between %
$(\tilde{g}_1,...,\tilde{g}_T) \sim \SM(D)$ and $(\tilde{g}'_1,...,\tilde{g}'_T) \sim \SM(D')$.
As mentioned, prior DP-SGD audits~\cite{jayaraman2019evaluating,jagielski2020auditing,nasr2021adversary,nasr2023tight} typically compute a natural ``score'' from each thresholded observation.
However, in the context of shuffling, there is no known score to distinguish between the outputs.

In this work, we draw from the Neyman-Pearson~\cite{neyman1933ix} lemma, which states that the optimal way to distinguish between two distributions is by thresholding the output of the \textit{likelihood ratio} function.
Specifically, the likelihood ratio function computes the ratio of probabilities that the outputs are from $\SM(D)$ or $\SM(D')$.
Here, the probability that $(\tilde{g}_1,...,\tilde{g}_T)$ is from $\SM(D)$ (or $\SM(D')$) can be split into $T$ cases depending on which batch the target record $+1$ (or $\bot$) appears in.
To that end, we calculate the likelihood ratio for the adjacent datasets $D = (+1, -1, ..., -1)$ and $D' = (\bot, -1, ..., -1)$ as follows (we let $\tilde{\mathbf{g}} = (\tilde{g}_1,...,\tilde{g}_T)$):
\begin{equation*}
\small
\begin{split}
  \Lambda(\tilde{\mathbf{g}}) = & \frac{\Pr[\tilde{\mathbf{g}} | \SM(D)]}{\Pr[\tilde{\mathbf{g}} | \SM(D')]}\\
   = & \frac{\sum_t T^{-1} \Pr[\tilde{\mathbf{g}} | \SM(D) \land +1\text{ is in batch } t]}{\sum_t T^{-1} \Pr[\tilde{\mathbf{g}} | \SM(D') \land \bot\text{ is in batch } t]} =  \\
  = & \frac{\sum_t \Pr[\tilde{g}_t | \mathcal{N}(-B + 2, \sigma^2)] \prod_{t' \neq t} \Pr[\tilde{g}_{t'} | \mathcal{N}(-B, \sigma^2)]}{\sum_t \Pr[\tilde{g}_t | \mathcal{N}(-B + 1, \sigma^2)] \prod_{t' \neq t} \Pr[\tilde{g}_{t'} | \mathcal{N}(-B, \sigma^2)]}
\end{split}
\end{equation*}

Note that for arbitrary adjacent datasets where all records are different, $\Lambda(\cdot)$ would be computationally intractable to compute since we would need to account for all possible permutations induced by shuffling.
However, since~\cite{chua2024private}'s conjecture assumes that all records except one are identical, this reduces the number of ``cases'' to be considered to $T$, thus making $\Lambda(\cdot)$ tractable.

\descr{Multiple Epochs.}
Although~\cite{chua2024private} only considers one set of batches (i.e., corresponding to 
a single ``epoch'' of training), we extend to multiple epochs as training private models typically involves %
multiple epochs.
We rely on the fact that the batch sampling and aggregation are done \textit{independently} across multiple epochs; thus, the likelihood across multiple epochs is the product of the likelihoods of each epoch, i.e.,
letting $\mathcal{G} = \begin{bmatrix} -\, \tilde{\mathbf{g}}^1 - \\ \vdots \\ -\, \tilde{\mathbf{g}}^E - \end{bmatrix}$, we define the likelihood ratio as:

{\small
\begin{align*}
&\Lambda^{\SM}(\mathcal{G})  = \prod_{i = 1}^E \Lambda(\tilde{\mathbf{g}}^i)= \\[-0.5ex] 
& = \prod_{i = 1}^E \frac{\sum_t \Pr[\tilde{g}^i_t | \mathcal{N}(-B + 2, \sigma^2)] \prod_{t' \neq t} \Pr[\tilde{g}^i_{t'} | \mathcal{N}(-B, \sigma^2)]}{\sum_t \Pr[\tilde{g}^i_t | \mathcal{N}(-B + 1, \sigma^2)] \prod_{t' \neq t} \Pr[\tilde{g}^i_{t'} | \mathcal{N}(-B, \sigma^2)]} \nonumber
\end{align*}
}
\indent Overall, our experimental evaluation (presented below) %
shows that auditing using likelihood ratios is very effective at estimating the privacy leakage from the \SMFull.
In this idealized setting, we observe a substantial gap (up to 4$\times$) between the privacy leakage observed empirically and the theoretical guarantees from the Poisson subsampling analysis (recall that there are no known guarantees for shuffling).
This further motivates investigating whether this gap is also observable %
in real-world settings using DP-SGD.

\begin{figure}[t]
  \centering
\hspace{-0.2cm}\includegraphics[width=1.02\linewidth]{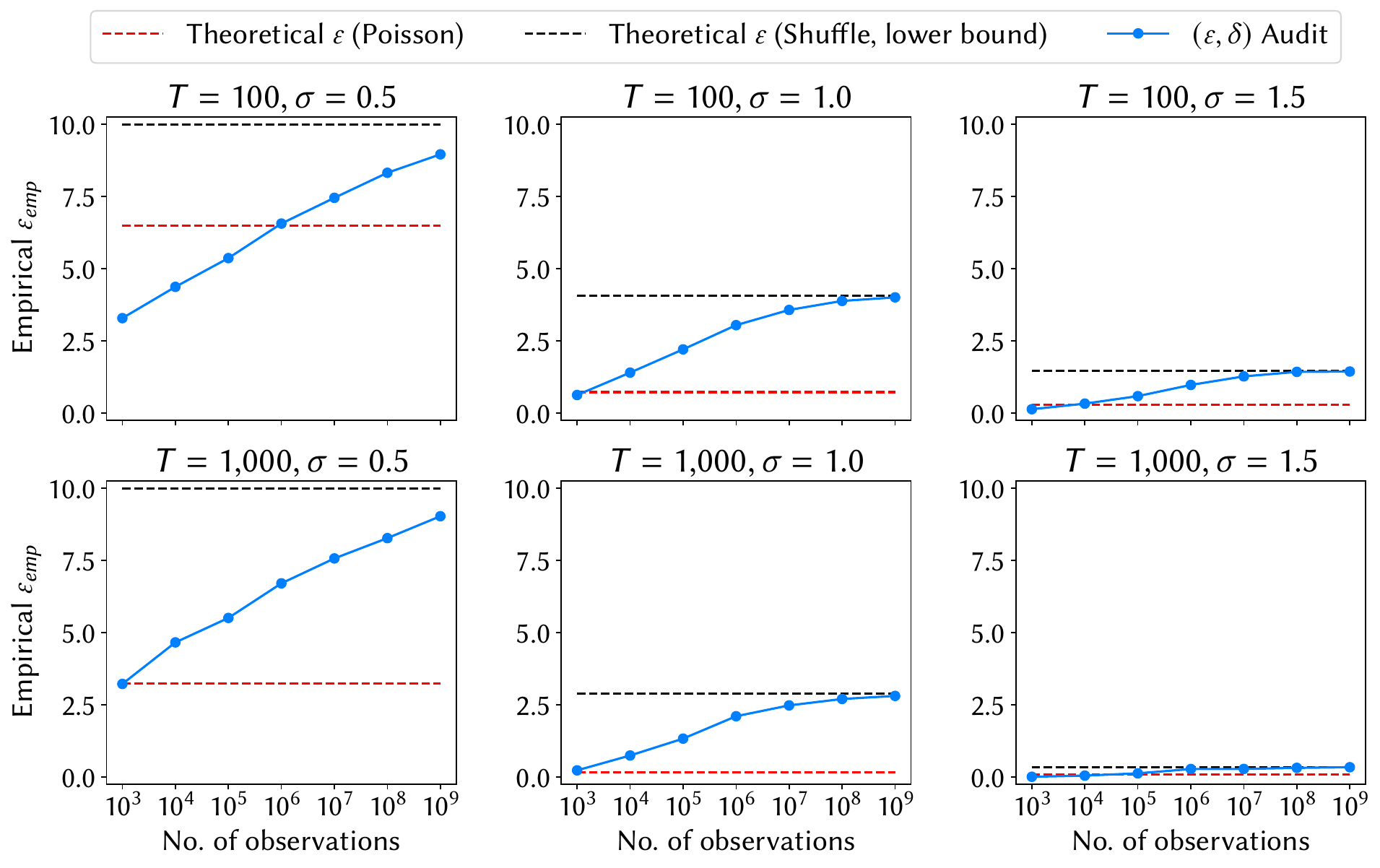}
  \caption{Auditing \SM for various number of steps $T$ and noise multipliers $\sigma$. Theoretical $\varepsilon$ (Poisson) represents the theoretical guarantees if Poisson subsampling had been used, while Theoretical $\varepsilon$ (Shuffle, lower bound) denotes the lower bounds from~\cite{chua2024private}.}
  \label{fig:param_sweep}
\end{figure}

\subsection{Experiments}\label{sec:results_audit_sm}

\subsubsection{Varying Batch Size and Noise Multiplier}

We start by fixing both the number of epochs and %
the batch size ($B$) to 1.
We vary the \textit{number of steps} for a single epoch, $T$, varying the dataset size, and can then compute the theoretical Poisson analysis $\varepsilon$ using the privacy loss random variable (PRV) accountant for a given $T$ and noise multiplier $\sigma$.
Furthermore, to assess computational complexity, we audit each setting over a varying number of game runs, each yielding a single observation.
For context, we also consider the lower bound theoretical $\varepsilon$ previously calculated by Chua et al.~\cite{chua2024private}.

In Figure~\ref{fig:param_sweep}, we plot the empirical $\empeps$ observed when auditing the \SMFull with the numbers of steps $T$ at 100 and 1,000 and noise multipliers $\sigma$  at $0.5, 1.0, 1.5$.
For several parameters, the empirical privacy leakage is substantially larger than the theoretical upper bounds from the Poisson analysis.
Specifically, with $T = 100$, we obtain $\empeps = 8.96, 4.01$, and $1.44$, respectively, for  $\sigma = 0.5, 1.0,1.5$ against theoretical upper bounds suggesting only $\varepsilon = 6.49, 0.73,0.30$ (over $10^9$ observations).

This gap varies with different numbers of steps taken in each epoch ($T$) and different noise multipliers ($\sigma$).
More precisely, it is substantially larger for larger $T$ (corresponding to small batch sizes) and smaller $\sigma$.
For instance, the gap between $\varepsilon$ and $\empeps$ is $5.78$ for $T = 1000$, $\sigma = 0.5$ vs.~$1.14$ for $T = 100$, $\sigma = 1.5$.
However, larger numbers of steps may not always result in larger empirical privacy leakage estimates either.
At $\sigma = 1.5$, for $T = 100$ and $1000$, we find $\empeps = 1.44$ and $\empeps = 0.34$, respectively.

Since the privacy loss distribution (PLD) used in~\cite{nasr2023tight} is not currently known for DP-SGD (Shuffle), recall that we are auditing using the $(\varepsilon, \delta)$-DP definition; as a result, we require a large number of observations for the $\empeps$ estimates to converge, especially when $\sigma$ is small.
For $\sigma \geq 1.0$, at $T = 100$, $\empeps$ converges after $10^8$ observations, but for $\sigma = 0.5$, it requires more than $10^9$ observations (we do not explore beyond $10^9$ due to computational constraints). 

As expected, given enough observations, our audits match but do not exceed the theoretical lower bounds calculated by Chua et al.~\cite{chua2024private}.
This not only confirms that our auditing procedure is effective but also that it might be possible to convert the \emph{lower-bound privacy analysis} from~\cite{chua2024private} into an \emph{upper-bound privacy guarantee}.
However, proving upper-bound privacy guarantees might not be trivial and is beyond our scope---thus, we leave it to future work.

\subsubsection{Exploring parameter settings}

Next, we audit %
using parameters from state-of-the-art differentially private models.
Specifically, we audit \SM in the settings %
used in~\cite{de2022unlocking} and~\cite{LTLH22} to train private image classification models and large language models, respectively\footnote{We audit the different parameter settings using our own shuffling implementation instead of auditing existing implementations both for computational efficiency and to freely vary the hyperparameters (e.g., batch size).}.
Note that De et al.~\cite{de2022unlocking} explicitly state they train their %
models with shuffling but report DP guarantees as though Poisson sampling was used.
Whereas, Li et al.~\cite{LTLH22} do not mention using shuffling, but a review of their codebase shows they use the default Pytorch implementation of \texttt{DataLoader}, which shuffles the dataset.

In total, we audit all 101 combinations of parameters (i.e., noise multiplier, $\sigma$, number of steps per epoch, $T$, and number of epochs $E$) extracted from~\cite{de2022unlocking} and \cite{LTLH22} using $10^7$ observations.
Overall, we find a substantial gap between the empirical privacy leakage under shuffling and the theoretical Poisson analysis in two-thirds of the settings studied.
In Table~\ref{tab:real_params}, we report the maximum gaps %
between $\empeps$ and $\varepsilon$,
over several datasets, for parameter settings used in prior work.
The largest gap (4.25$\times$) occurs with large language models, probably owing to the small noise multiplier used to optimize utility.

Overall, these experiments indicate that, under the \SMFull's ideal conditions, training with shuffling instead of Poisson subsampling incurs substantial privacy leakage, also in settings commonly used for training state-of-the-art models.
This further motivates assessing the validity of the DP guarantees reported by state-of-the-art models.

\begin{table}[!t]
\vspace{0.2cm}
  \centering
  \small
    \setlength{\tabcolsep}{4pt}
  \begin{tabular}{@{}c@{}l@{}r|rrr|rrr}
  \toprule
  {\bf Paper~} & {\bf Dataset} & {\bf Size} & $\sigma$ & $T$ & $E$ & $\varepsilon$ & $\empeps$ & Gap\\
  \midrule
  \cite{de2022unlocking}~ & Places-365~ & 1.8M & 1.00 & 440 & 509 & 7.53 & 13.54 & 1.80$\times$ \\
  \cite{de2022unlocking}~ & CIFAR-10~ & 60K & 3.00 & 11 & 168 & 6.24 & 6.39 & 1.02$\times$\\
  \cite{LTLH22}~ & SST-2~ & 60K & 0.79 & 117 & 10 & 3.00 & 9.80 & 3.27$\times$\\
  \cite{LTLH22}~ & QNLI~ & 100K & 0.87 & 195 & 30 & 3.00 & 10.60 & 3.53$\times$\\
  \cite{LTLH22}~ & MNLI~ & 400K & 0.73 & 781 & 50 & 3.00 & 12.73 & 4.25$\times$\\
  \cite{LTLH22}~ & Persona-Chat~ & 400K & 0.82 & 254 & 30 & 2.99 & 11.08 & 3.70$\times$\\
  \bottomrule
  \end{tabular}
\sptable  
  \caption{Parameter settings used in prior work where the empirical privacy leakage ($\empeps$) observed from shuffling is appreciably larger than the theoretical $\varepsilon$ from Poisson sampling.}%
  \label{tab:real_params}
\end{table}

\subsection{Takeaways}
As mentioned earlier, we experiment with \SMFull and study the impact of various parameter settings (noise scale, batch size, and number of epochs) on the empirical privacy leakage observed.
Our experiments show that the empirical privacy leakage from shuffling can be substantially larger than the theoretical guarantees given by the Poisson subsampling analysis.

Crucially, this gap is prominent in the parameter settings used to train SOTA models in prior work~\cite{de2022unlocking,LTLH22} as well.
For instance, De et al.~\cite{de2022unlocking} reportedly use shuffling to fine-tune an NF-ResNet-50 model on the Places-365 dataset at a theoretical $\varepsilon = 7.53$, but our audits show that the actual privacy leakage is almost double ($\empeps = 13.59$).

\section{Auditing DP-SGD (Shuffle)}
\label{sec:audit_dpsgd}

In this section, we present our novel auditing procedure %
for DP-SGD (Shuffle). %
We start by defining various threat models specific to the shuffling setting and then present our algorithm.

\subsection{Adversarial Modeling}
\label{sec:threat_models}

Although DP-SGD (Shuffle)'s structure is very similar to that of \SM, there are practical considerations that can affect the empirical privacy estimates. %
For instance, since the inputs to the mechanism are directly aggregated and all intermediate noisy aggregates are released, adversaries for the \SM are, by default, given as much power as the theoretical worst-case DP adversary.
By contrast, DP-SGD first calculates the gradient of each input before aggregating them and may only output the final trained model, thus making it unclear which aspects of the mechanism the adversary is given access to.

\descr{Auditing with a \WC Adversary.}
We start by considering the worst-case adversary, assuming they can insert the gradients of all records in the dataset.
Although this may %
destroy model utility, recall that the main goal of DP auditing is to verify that the DP bounds are correct, and DP is a worst-case guarantee that should hold against the most powerful adversary. %
In fact, the privacy analysis of DP-SGD assumes that the adversary can indeed insert arbitrary gradients for all records~\cite{abadi2016deep}.

Overall, adversaries used in the DP auditing literature are usually given strong capabilities, including, e.g., (active) white-box access to the model~\cite{nasr2021adversary,nasr2023tight}.
More precisely, DP-SGD audits~\cite{nasr2021adversary,nasr2023tight,annamalai2024what,cebere2024tighter,pillutla2024unleashing,steinke2024privacy,mahloujifar2024auditing,xiang2025privacy} typically consider a ``white-box with gradient canaries'' setting where not only can the adversary choose the target record $(\hat{x}, \hat{y})$, but they also have access to the model parameters at each update step and can arbitrarily insert gradients of samples (i.e., gradient canary).
Our work also operates in this setting.

\descr{Auditing with a \Nat Adversary.} %
Prior DP-SGD audits~\cite{nasr2021adversary,nasr2023tight,annamalai2024what,cebere2024tighter,pillutla2024unleashing,steinke2024privacy,mahloujifar2024auditing,xiang2025privacy} have also experimented with adversaries that can only insert the gradient of the target record, obtaining relatively tight audits for DP-SGD (Poisson).
In this setting, all records except the target have natural gradients (i.e., not adversarially crafted). 
While we expect models audited using these adversaries to have the same utility as production models (the canary gradient has minimal impact on it),
our experiments show the resulting audits to be relatively loose. %
As a result, we only report experimental results in the context of analyzing the impact of varying adversarial capabilities on the tightness of the audit (as also done in prior work~\cite{jagielski2020auditing,nasr2021adversary,nasr2023tight,annamalai2024nearly,steinke2024privacy}). %

\descr{Auditing with a \PI Adversary.} %
Theoretically, the worst-case adversary assumed by the DP guarantees has access to all aspects of the algorithm not involved in the randomness for noise addition and batch selection, i.e., it can modify the gradients of \emph{all} samples.
Nevertheless, we believe it is interesting and useful to relax this assumption %
and introduce the so-called \PI adversary, which can only modify \emph{selected} samples' gradients.
Specifically, this adversary only inserts the gradients of the target record and the final record \textit{in each batch}.
We are motivated to do so as one of the reasons why \Nat audits are loose could stem from the \textit{bias} introduced by the other samples in each batch, which our likelihood-based auditing procedure could be sensitive to.
Therefore, by leaving the remaining $B - 1$ samples in each batch untouched, we can precisely study the impact of the \textit{bias} introduced by other samples on our auditing procedure.
Similar to the \WC, the adversary can also insert opposing gradients, 
thus, making the batches containing the target or zero-out record more obvious.
Overall, models audited under this threat model should still maintain similar utility as production models, while we expect them to result in tighter audits.

\descr{Auditing with Strong Adversaries.}
Although our adversarial models consider relatively strong, gradient-crafting adversaries, arguably, this does not diminish the value and impact of our analysis.
First of all, as we discuss in Section~\ref{sec:related}, prior work has considered similarly strong adversaries: for instance, in the context of DP-SGD (Poisson), Nasr et al.~\cite{nasr2021adversary} rely on adversaries that can craft arbitrarily malicious datasets (i.e., ``Pathological Dataset''). 
In other words, our Worst-Case adversary can be modeled with the Pathological Dataset adversary; thus, our threat model does not involve adversaries stronger than those used in prior auditing work (see Appendix~\ref{app:worst_case}).

Overall, it is common to instantiate audits with different adversarial models and evaluate the resulting impact on their tightness~\cite{nasr2021adversary,nasr2023tight,annamalai2024what}.
We follow a similar strategy and consider several adversarial models that are specifically catered to DP-SGD (Shuffle); this also allows us to evaluate the impact of adversarial capabilities on different aspects of DP-SGD (Shuffle) algorithms in terms of privacy leakage.

Perhaps more importantly, being a worst-case definition, DP is meant to provide formal guarantees that hold also against worst-case adversaries.
One of the main goals of DP auditing is to verify that empirical leakage does not exceed these (theoretical) DP guarantees.
In our case, in particular, the main research question we focus on is whether or not privacy discrepancies emerge from designing DP-SGD algorithms that use shuffling but reporting guarantees as if Poisson subsampling was used.
Therefore, the adversaries we experiment with should be seen as part of our auditing toolkit, more than in the context of how ``realistic'' it would be to instantiate them.

\subsection{Auditing Procedure}

\begin{algorithm}[t]
    \small
    \caption{The DP-SGD algorithm with the modifications we make to audit DP-SGD (Shuffle) with different adversaries. Modifications for \Nat, \PI, and \WC  are reported in \textcolor{darkred}{red}. Additional modifications for \PI and \WC only are in \textcolor{blue}{blue} and for \WC only in \textcolor{orange}{orange}.}\label{alg:dpsgd_audit}
    \begin{algorithmic}[1]
    \Require Dataset, $D$. Epochs, $E$. Batch Size, $B$. Learning rate, $\eta$. Batch sampler, $\mathcal{B}$. Loss function, $\ell$. Initial model parameters, $\theta_0$. Noise multiplier, $\sigma$. Clipping norm, $C$. \textcolor{darkred}{Target record, $(\hat{x}, \hat{y})$}. \textcolor{darkred}{Canary gradient, $\hat{g}$}. \textcolor{darkred}{Zero-out record, $(x_\bot, y_\bot)$}.
    \State $T \leftarrow |D| / B$
    \For{$i \in [E]$}
      \State $\theta^i_1 \leftarrow \theta^{i - 1}_T$
      \State Sample batches $B_1, ..., B_T \leftarrow \mathcal{B}(D, B)$
      \For{$t \in [T]$}
        \For{$(x_j, y_j) \in B_t$}
          \color{darkred}
          \If{$(x_j, y_j) = (x_\bot, y_\bot)$}
            \State $g_j \leftarrow \mathbf{0}$
          \ElsIf{$(x_j, y_j) = (\hat{x}, \hat{y})$}
            \State $g_j \leftarrow \hat{g}$
          \color{blue}
          \ElsIf{$j = B - 1 \land (\hat{x}, \hat{y}), (x_\bot, y_\bot) \notin B_t$}
            \State $g_j \leftarrow -\hat{g}$
          \color{black}
          \Else
            \State $g_j \leftarrow \nabla \ell((x_j, y_j); \theta^i_t)$
            \color{orange}
            \State $g_j \leftarrow 0$
            \color{black}
          \EndIf
          \State $\bar{g}_j \leftarrow g_j / \max\left(1, \frac{||g_j||_2}{C}\right)$
        \EndFor
        \State $\tilde{g} \leftarrow \frac{1}{B} \left(\sum_j \bar{g}_j + \mathcal{N}(0, C^2\sigma^2\mathbb{I})\right)$
        \State $\theta^i_{t + 1} \leftarrow \theta^i_t - \eta\tilde{g}$
        \color{darkred}
        \State $o^i_t \leftarrow \langle\tilde{g}, \hat{g}\rangle$
        \color{black}
      \EndFor
    \EndFor
    \color{darkred}
    \State \Return $\mathcal{O} = \begin{bmatrix} 
      o^1_1 & \dots  & o^1_T\\
      \vdots & \ddots & \vdots\\
      o^E_1 & \dots  & o^E_T
      \end{bmatrix}$
    \color{black}
    \end{algorithmic}
\end{algorithm}

In Algorithm~\ref{alg:dpsgd_audit}, we outline the DP-SGD algorithm~\cite{abadi2016deep} with the modifications needed to audit with \Nat, \PI, and \WC adversaries.
Lines 7-8 correspond to enforcing that the $\bot$ record has a $\mathbf{0}$ gradient, which is needed for auditing with zero-out adjacent datasets.

Then, lines 9-10 are standard steps for active white-box adversaries that insert the gradient of the target record.
Lines 11-12 correspond to inserting the gradient of the final record in each batch, which only occurs with \PI and \WC (but not \Nat). 
Specifically, the adversary inserts the canary gradient in the opposite direction for each batch that does not contain both the target and zero-out records.
This closely resembles the conjectured worst-case datasets by Chua et al.~\cite{chua2024private} as the target record's gradient has an equivalent magnitude but opposite direction to the final records in all other batches, which makes audits in this model tighter.
This does not break or change the underlying DP guarantees of DP-SGD (Shuffle) as this step is applied regardless of whether the target or zero-out record is present in the dataset and thus applies to $D$ and $D'$ equally.

In line 15, the \WC adversary additionally zeroes the gradients of all other samples, effectively removing the bias from the other samples.
Finally, all adversaries compute the dot product between the privatized and canary gradient as ``outputs'' (line 21) and release the full matrix of ``outputs'' across all batches and epochs (line 24).
Similar to \SM, the adversary computes and thresholds the likelihood ratio to calculate empirical privacy leakage estimates.

In our auditing procedure, the adversary does not have access to the underlying batch sampler $\mathcal{B}$, but only to the output matrix $\mathcal{O}$.
We do so as we focus on determining the impact of alternate subsampling schemes on the empirical privacy leakage. %
This allows us not only to assess whether the adversary can leak more privacy without knowing the specifications of the batch sampler but also to detect bugs within the shuffling implementations (see Section~\ref{sec:debug_shuffle}).

Overall, our auditing procedure assumes an adversary that can access all parts of training other than the clipping, noise addition, and batch selection steps and is only slightly weaker than the adversary assumed by DP-SGD's privacy analysis (i.e., the optimal adversary can also choose a pathological dataset $D$ and has access to the batch sampler).

\subsection{Computing Likelihood Ratios}
The last step in the audit is %
for the adversary to calculate the score assigned to the mechanism's output using the Neyman-Pearson lemma.
While this is similar to the \SM setting, the likelihood ratios differ depending on the threat model.
Mainly, $o^i_t = \langle \tilde{g}, \hat{g} \rangle \approx 0$ if $\hat{g}$ or $-\hat{g}$ were not inserted by the adversary in Steps 10 or 12 and $o^i_t \approx +1$ and $\approx -1$, respectively, otherwise.
For each threat model, we calculate $\Lambda(\mathcal{O})$  as follows.\smallskip

\begin{itemize}[leftmargin=10pt]
\item \Nat ($\Lambda^{\text{TC}}$):
{\small
\begin{equation*}
  \prod_{i = 1}^E \frac{\sum_t \Pr[o^i_t | \mathcal{N}(+1, \sigma^2)] \prod_{t' \neq t} \Pr[o^i_{t'} | \mathcal{N}(0, \sigma^2)]}{\prod_{t} \Pr[o^i_t | \mathcal{N}(0, \sigma^2)]}
\end{equation*}}

\item \PI ($\Lambda^{\text{PI}}$):
{\small
\begin{equation*}
  \prod_{i = 1}^E \frac{\sum_t \Pr[o^i_t | \mathcal{N}(+1, \sigma^2)] \prod_{t' \neq t} \Pr[o^i_{t'} | \mathcal{N}(-1, \sigma^2)]}{\sum_t \Pr[o^i_t | \mathcal{N}(0, \sigma^2)] \prod_{t' \neq t} \Pr[o^i_{t'} | \mathcal{N}(-1, \sigma^2)]}
\end{equation*}}

\item \WC ($\Lambda^{\text{WC}}$):
{\small
\begin{equation*}
  \prod_{i = 1}^E \frac{\sum_t \Pr[o^i_t | \mathcal{N}(-B + 2, \sigma^2)] \prod_{t' \neq t} \Pr[o^i_{t'} | \mathcal{N}(-B, \sigma^2)]}{\sum_t \Pr[o^i_t | \mathcal{N}(-B + 1, \sigma^2)] \prod_{t' \neq t} \Pr[o^i_{t'} | \mathcal{N}(-B, \sigma^2)]}\\[1ex]
\end{equation*}}

\end{itemize}

\section{Experimental Evaluation}
We now present an experimental evaluation of our auditing techniques, aiming to compare the empirical privacy leakage observed from mechanisms that use shuffling ($\empeps$) with the upper-bound privacy leakage guaranteed by the analysis using Poisson subsampling ($\epsilon$).

\subsection{Experimental Overview}

\descr{Datasets.}
We use three datasets commonly used in the DP auditing literature: FMNIST~\cite{xiao2017fashion}, CIFAR-10~\cite{krizhevsky2009learning}, and Purchase-100 (P100)~\cite{shokri2017membership}.
Due to computational constraints, we only take samples corresponding to two labels from FMNIST and CIFAR-10 and downsample all datasets only to include 10,000 samples, as also done in prior work~\cite{jagielski2020auditing}. 
Therefore, our FMNIST dataset contains 10,000 28x28 grayscale images from one of two classes (`T-shirt' and `Trouser'), CIFAR-10 contains 10,000 3x32x32 RGB images from one of two classes (`Airplane' and `Automobile'), and P100 consists of 10,000 records with 600 binary features from one of 100 classes.

\descr{Models.}
For CIFAR-10, we train a moderate-sized Convolutional Neural Network (CNN) drawn from prior work~\cite{dormann2021not}.
For FMNIST and P100, we train a small LeNet and an MLP model, respectively.
We refer to Appendix~\ref{app:model_arch} for more details on the model architectures.

\descr{Experimental Testbed.}
We run all experiments on a cluster using 4 NVIDIA A100 GPUs, 64 CPU cores, and 100GB RAM.
Auditing a single model using $10^6$ observations took 38.8, 17.1, and 5.70 hours for the shallow CNN, LeNet, and MLP models, respectively.

\descr{Metrics and Parameters.} For all experiments, we report lower bounds with 95\% confidence (Clopper-Pearson~\cite{clopper1934use}) along with the mean and standard deviation values of $\empeps$ over five independent runs.
For simplicity, we set $\delta = 10^{-5}$, gradient clipping norm to $C = 1.0$, and choose the learning rate $\eta$ by hyperparameter tuning from a logarithmic scale.

\subsection{Auditing DP-SGD (Shuffle) with \PINorm}\label{sec:results_audit_dpsgd}

Next, we audit DP-SGD (Shuffle) in the \PI model described in Section~\ref{sec:threat_models}. %
We do so because we expect \PI to produce tighter audits than \Nat, without destroying model utility like for \WC.

In all experiments, we use a randomly sampled gradient from the unit ball as the canary gradient, although we did not notice any significant difference between different types of canary gradients (e.g., dirac) in preliminary experiments.

\subsubsection{Comparing trade-off curves}
\label{sec:compare_tradeoff}

\begin{figure}[t]
  \centering
  \includegraphics[width=\mywidth\linewidth]{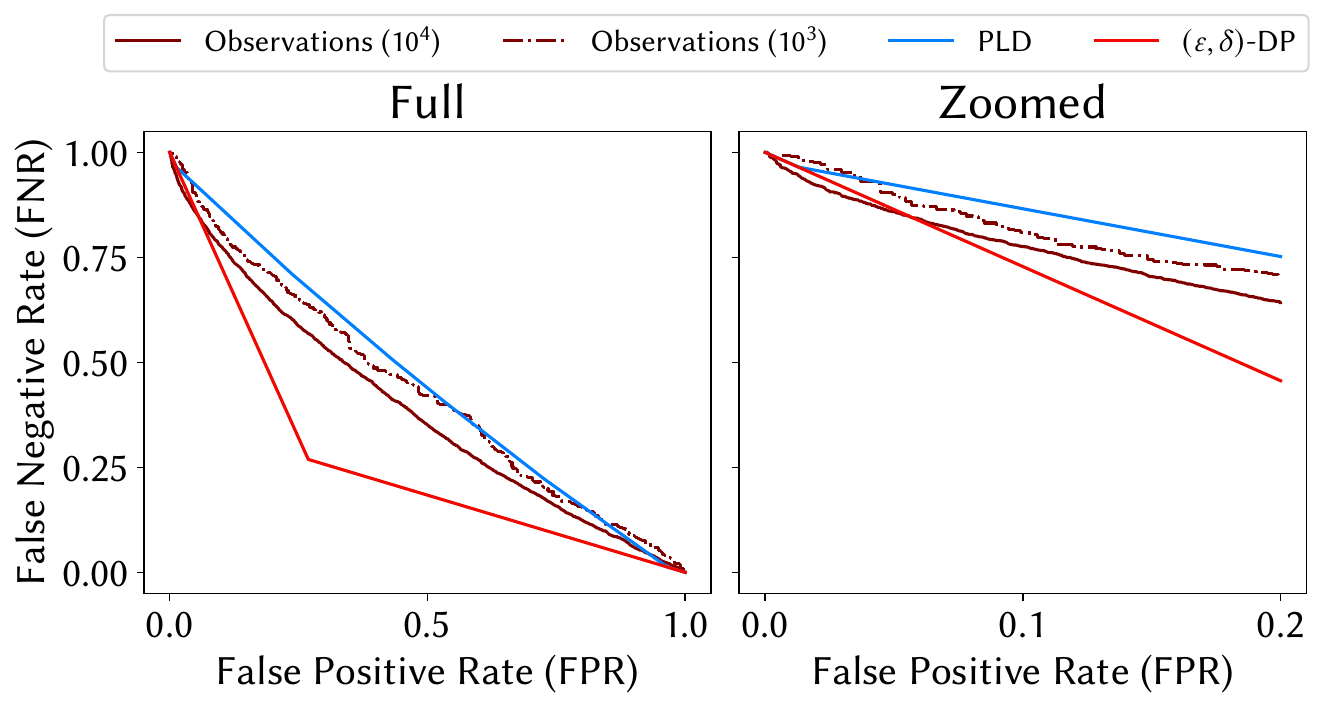}
  \caption{Comparison of tradeoff curves for auditing DP-SGD (Shuffle) (95\% Clopper-Pearson upper bound) at theoretical $\varepsilon = 1.0$ vs.~the corresponding theoretical tradeoff curves from \approxdp and from the PLD analysis of DP-SGD (Poisson). The plot on the right is zoomed in on $0 \leq$ FPR $\leq 0.2$.}
  \label{fig:compare_tradeoffs}
\end{figure}

We start by auditing a CNN model on CIFAR-10 for one epoch using shuffling with batch size $B = 100$, calibrating the noise multiplier to satisfy $\varepsilon = 1.0$ if Poisson subsampling was used.
In Figure~\ref{fig:compare_tradeoffs}, we plot the FPR-FNR curve (95\% Clopper-Pearson upper bound) from training $10^3$ and $10^4$ models and the curves for the corresponding theoretical $(\epsilon,\delta)$-DP at $\epsilon=1.0$
and PLD bounds expected if Poisson subsampling had been used.
We note that an \emph{upper bound} on FPR and FNR is used to compute the corresponding \emph{lower bound} on empirical $\empeps$ as explained previously in Section~\ref{sec:estimate_eps}.
Due to computational constraints, we are only able to train $10^4$ models for this experiment.

Overall, we find substantial gaps %
regardless of the number of observations.
More precisely, even a relatively small number of observations ($10^3$) is enough to detect that the DP-SGD (Shuffle) implementation violates the theoretical privacy guarantees provided by the Poisson subsampling analysis.
However, while auditing with the PLD curve would allow an adversary to identify that a given implementation of DP-SGD did not specifically use Poisson subsampling, it may not necessarily mean that the underlying implementation does not satisfy %
$(\epsilon,\delta)$-DP in general.

To estimate the empirical privacy leakage observed from DP-SGD (Shuffle), we would need to audit using the $(\varepsilon, \delta)$-DP definition instead.
In Figure~\ref{fig:compare_tradeoffs}, this would correspond to comparing the observed FPR-FNR curves with the $(\epsilon,\delta)$-DP %
curve instead ($\epsilon=1.0$).
Here, there is a much smaller gap, with the trade-off curve for $10^4$ observations only violating $(\epsilon,\delta)$-DP %
at FPRs $< 0.05$, and the trade-off curve for $10^3$ not %
violating the theoretical DP guarantees at all.
Therefore, $10^3$ observations are not enough to detect a violation of $(\epsilon,\delta)$-DP %
guarantee: the violation can only be detected by using at least $10^4$ observations at very low FPRs.%

\begin{figure}[t]
  \centering
  \includegraphics[width=\mywidth\linewidth]{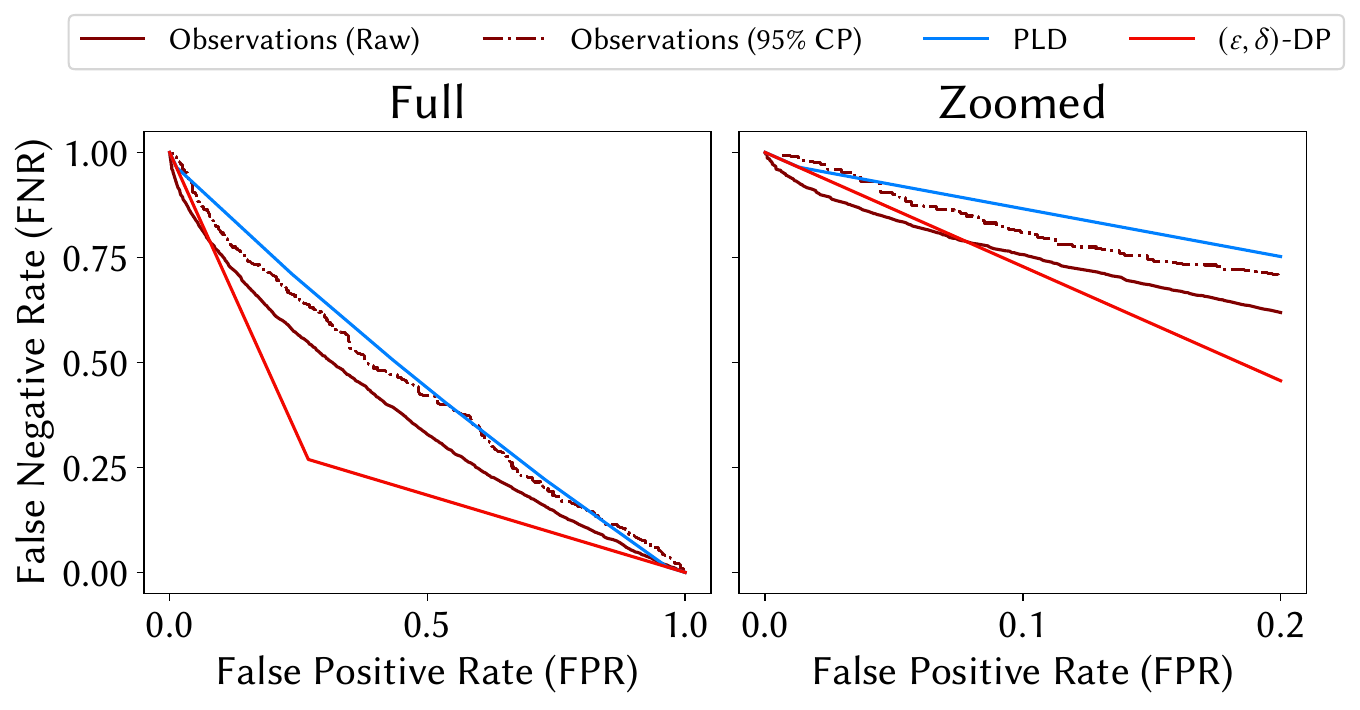}
  \caption{Comparison of tradeoff curves for %
  auditing DP-SGD (Shuffle) (Raw and 95\% CP upper bound, $10^3$ observations) vs.~the theoretical tradeoff curves from \approxdp and PLD analysis for DP-SGD (Poisson). The plot on the right is zoomed in on $0 \leq$ FPR $\leq 0.2$.}
  \label{fig:compare_tradeoffs_ci}
\end{figure}

\descr{Impact of CIs.} To better understand the dependence on the number of observations, we also analyze the impact of using Clopper-Pearson (CP) CIs.
In Figure~\ref{fig:compare_tradeoffs_ci}, we plot the trade-off curves from the same experiment with $10^3$ observations with and without the CP upper bounds.
We notice a large gap between the ``raw'' observed trade-off curve (w/o CP upper bounds) and the trade-off curve with CP upper bounds.
Specifically, when using the trade-off curve with CP bounds, we estimate a lower bound $\empeps$, which is always numerically smaller than the $\empeps$ estimated from the ``raw'' trade-off curve, thus making it more difficult to detect violations of \approxdp.
Nevertheless, in DP auditing, we are interested not only in estimating the empirical privacy leakage but also in the associated confidence level.
This allows us to reduce false positives and provide assurances that the privacy violations are real, especially when debugging implementations (as we do later in Section~\ref{sec:debug_shuffle}). %
As a result, we believe it necessary to use CP intervals in our auditing experiments, even though that requires more training runs.

\subsubsection{Varying batch size $B$ and theoretical $\varepsilon$}

\begin{figure}[t]
  \centering
  \includegraphics[width=\mywidth\linewidth]{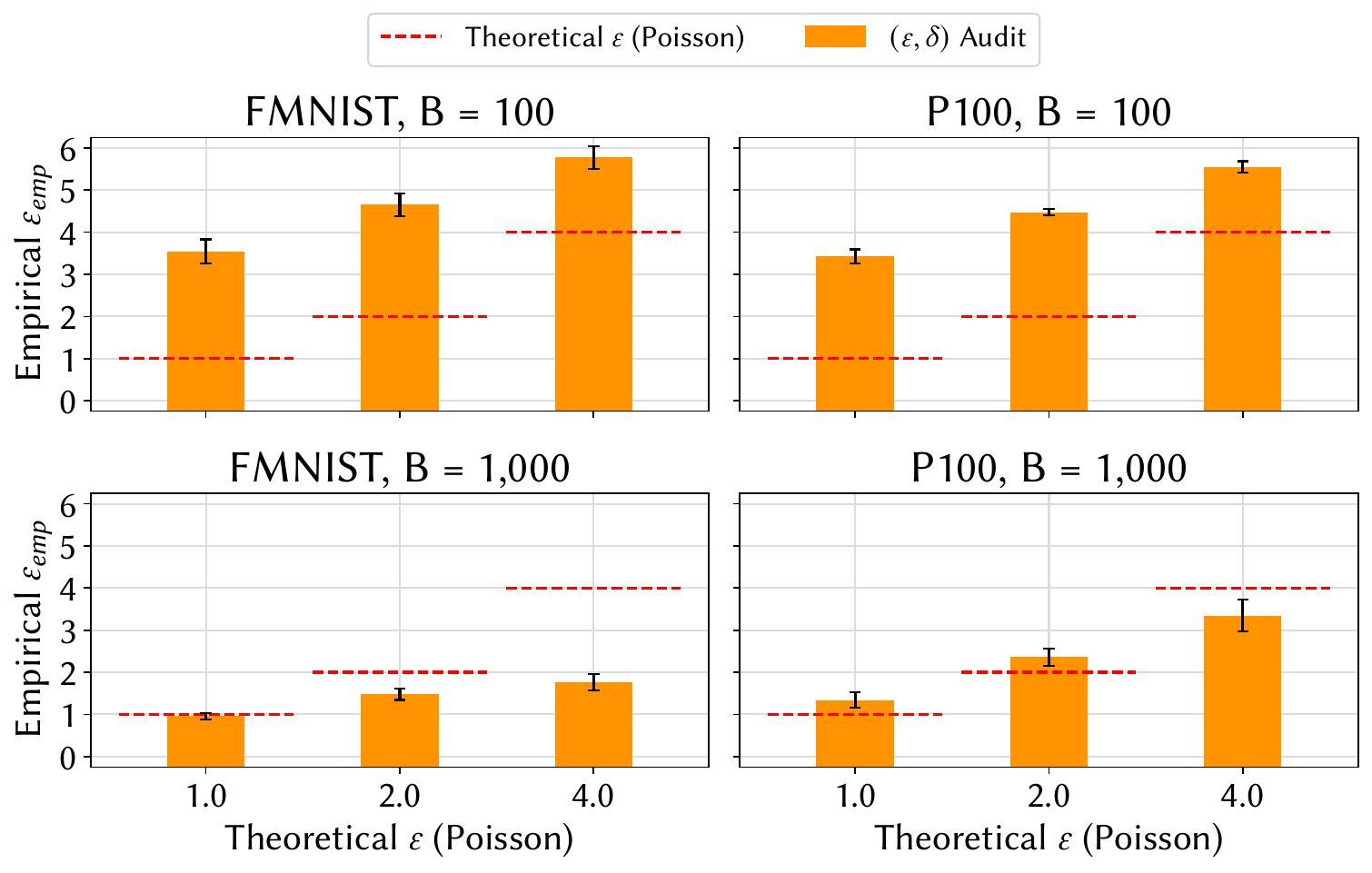}
  \caption{Auditing DP-SGD (Shuffle) at various batch sizes ($B$) and privacy levels ($\varepsilon$).}
  \label{fig:batch_exp}
\end{figure}

Next, we evaluate the impact of the batch size and theoretical (Poisson) $\varepsilon$ on the empirical privacy leakage observed from DP-SGD (Shuffle). %
Due to computational constraints, rather than a CNN on CIFAR-10, we audit a LeNet model on FMNIST and an MLP model on P100 over $10^6$ observations.
In Figure~\ref{fig:batch_exp}, we plot the empirical leakage estimates $\empeps$ observed from auditing DP-SGD (Shuffle) at various batch sizes and privacy levels $\varepsilon$.

With  $B = 100$, the empirical privacy leakage is significantly larger than the theoretical guarantees across both datasets and privacy levels.
Specifically, for %
$\varepsilon = 1.0, 2.0, 4.0$, privacy leakage from the models trained on FMNIST and P100 amount to, respectively, $\empeps = 3.46, 4.66, 5.78$ and $3.42, 4.48, 5.55$.

However, with a larger batch $B = 1000$, we only detect small gaps %
in some settings.
Under \PI, the gradients of all samples remain unchanged except for the target record and final record in each batch.
Thus, this likely introduces a ``bias'' term when evaluating the output from each step, i.e., $o^i_t \leftarrow \langle \tilde{g}, \hat{g} \rangle$.
Although Nasr et al.~\cite{nasr2023tight} show that this bias does not affect audits of DP-SGD (Poisson), they simply threshold the output from each step.
Our auditing procedure is more complicated as we compute a likelihood ratio function across multiple steps, which we believe is more susceptible to the bias term, resulting in weaker audits than expected from \SM.
However, in several settings, the empirical privacy leakage is close to or already exceeds the theoretical Poisson guarantee, suggesting that given enough observations ($\approx 10^9$), we can potentially detect larger. %
Due to computational constraints, %
we leave this to future work.

\subsection{Varying Adversarial Capabilities}
\label{sec:diff_threat_models}
Our experiments thus far show that auditing DP-SGD (Shuffle) under \PI reveals large gaps between the empirical privacy leakage observed and the theoretical guarantees promised by the Poisson subsampling analysis. 
Next, we set out to evaluate the impact of different adversarial capabilities on the empirical privacy leakage.

\begin{figure}[t]
  \centering
  \includegraphics[width=\mywidth\linewidth]{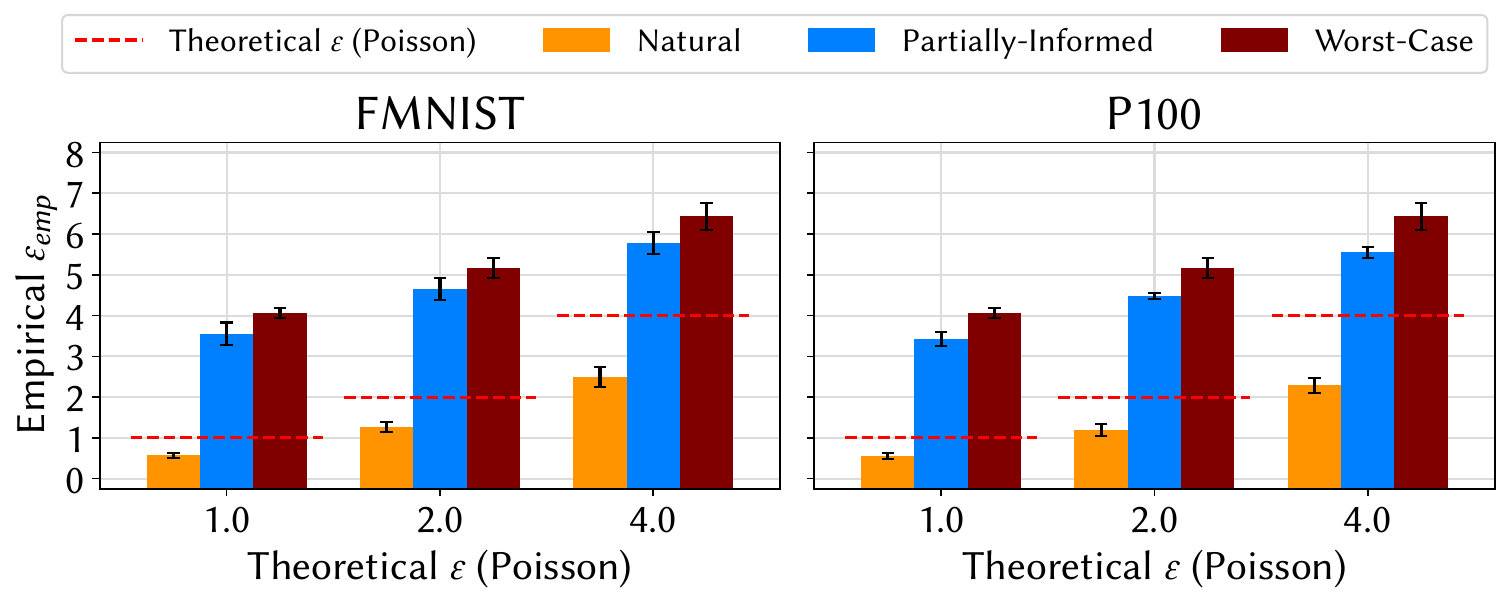}
  \caption{Auditing DP-SGD (Shuffle) under various threat models.} 
  \label{fig:threat_exp}
\end{figure}

In Figure~\ref{fig:threat_exp}, we report the different values of $\empeps$ when auditing DP-SGD (Shuffle) with the three different adversaries considered (see Section~\ref{sec:audit_dpsgd}). %
We fix the batch size to $B = 100$ and use $10^6$ observations.
Despite the ``bias'' from other training samples discussed above, the empirical privacy leakage under \PI is almost the same as that from the idealized \WC threat model (which corresponds to auditing \SM) at small batch sizes.
Specifically, on FMNIST, at $\varepsilon = 1.0, 2.0, 4.0$, auditing under \PI and \WC yield privacy leakage estimates, respectively, of $\empeps = 3.55, 4.66, 5.78$ and $\empeps = 4.06, 5.18, 6.43$.

However, estimates under \Nat are significantly weaker than the other two threat models.
In all settings, $\empeps$ under \Nat is actually smaller than the theoretical $\varepsilon$ guaranteed by Poisson subsampling, even though shuffling was used.
This is because the negative canary gradients inserted by the \PI adversary are crucial in ``identifying'' the batch containing the target (or zero-out) record in each epoch.
Specifically,  in the \PI threat model, without any noise, the adversary aims to distinguish between $(-1, ..., +1, ..., -1)$ and $(-1, ..., 0, ..., -1)$, which has a Hamming distance of 2 with high probability (assuming the `+1' and `0' appear in different batches).
Whereas under \Nat, they have to distinguish between $(0, ..., +1, ..., 0)$ and $(0, ..., 0)$, which only has a Hamming distance of 1.
Informally, the distributions of outputs observed under \Nat are less distinguishable than under \PI, thus resulting in lower empirical privacy leakage estimates.

\subsection{Takeaways}
Our experiments confirmed that the empirical privacy leakage observed from shuffling is much larger than the theoretial privacy leakage expected from the Poisson subsampling analysis when auditing models trained using the DP-SGD (Shuffle) algorithm.
Specifically, for a LeNet-5 classifier trained on FMNIST at theoretical $\varepsilon = 1.0$ in the %
\PI model, we find an empirical privacy leakage amounting to $\empeps = 3.46$, almost $3.5\times$ the theoretical guarantee.

We also show that audits using a weaker adversary, i.e., \Nat, the $\empeps$ is well below the expected theoretical guarantees $\varepsilon$, which confirms that strong adversarial models are necessary to audit shuffling mechanisms effectively.

\section{Debugging Shuffle Implementations}
\label{sec:debug_shuffle}

Implementing DP algorithms correctly is known to be challenging~\cite{opacusbug,prngreuse}, and DP violations have been found in the wild%
~\cite{tramer2022debugging,annamalai2024what}.
Specific to shuffling, Ponomareva et al.~\cite{ponomareva2023dp} report that for computational reasons, in many cases, datasets may not even be shuffled fully but only within a small buffer.
Therefore, bugs or variations to the shuffling procedure itself may be present in DP-SGD implementations, and, naturally, this would substantially affect their privacy leakage.

Searching public GitHub repositories (see Section~\ref{sec:bts}), we identify cases where the dataset is first batched, and the batches are then shuffled, rather than shuffling the samples before batching.
As visualized in Section~\ref{sec:bts}, this makes the batch with the target record substantially more identifiable in cases where the samples surrounding it are different from samples ``far away'' from it (in this setting, batching is done locally).
While these variations to the shuffling procedure intuitively affect the privacy analysis, in this section, we verify whether our auditing procedure can identify them and determine the scale of their impact on privacy leakage.
\subsection{Partial Shuffling}

\begin{algorithm}[t]
    \small
    \caption{Auditing Partial Shuffling (with Buffer $K$)}\label{alg:audit_partial_shuffle}
    \begin{algorithmic}[1]
    \Require Dataset, $D$. Number of observations, $N$. Target record, $(\hat{x}, \hat{y})$. Zero-out record, $(x_\bot, y_\bot)$. Significance level, $\alpha$. Privacy parameter, $\delta$.
    \For{$i \in [\frac{N}{2}]$}
      \State $\Theta[i] \leftarrow \text{DP-SGD}^{\text{PI}}_K(\{(\hat{x}, \hat{y})\} \cup D;-)$
      \State $\Theta'[i] \leftarrow \text{DP-SGD}^{\text{PI}}_K(\{(x_\bot, y_\bot)\} \cup D;-)$
    \EndFor

    \For{$k \in \{1, 10, 20, ..., 100\}$}
      \For{$i \in [\frac{N}{2}]$}
        \State $[\mathbf{o}_1 | ... | \mathbf{o}_T] \leftarrow \Theta[i]$
        \State $[\mathbf{o'}_1 | ... | \mathbf{o'}_T] \leftarrow \Theta'[i]$

        \State $\mathcal{S}[i] \leftarrow \Lambda^{\text{PI}}([\mathbf{o}_1 | ... | \mathbf{o}_k])$
        \State $\mathcal{S}'[i] \leftarrow \Lambda^{\text{PI}}([\mathbf{o'}_1 | ... | \mathbf{o'}_k])$
      \EndFor
      \State $\empeps[k] \leftarrow \text{EstimateEps}(\mathcal{S}, \mathcal{S}', \alpha, \delta)$
    \EndFor

    \State \Return $\max_k \empeps[k]$.
    \end{algorithmic}
\end{algorithm}

\begin{figure}[t]
  \centering
  \includegraphics[width=\linewidth]{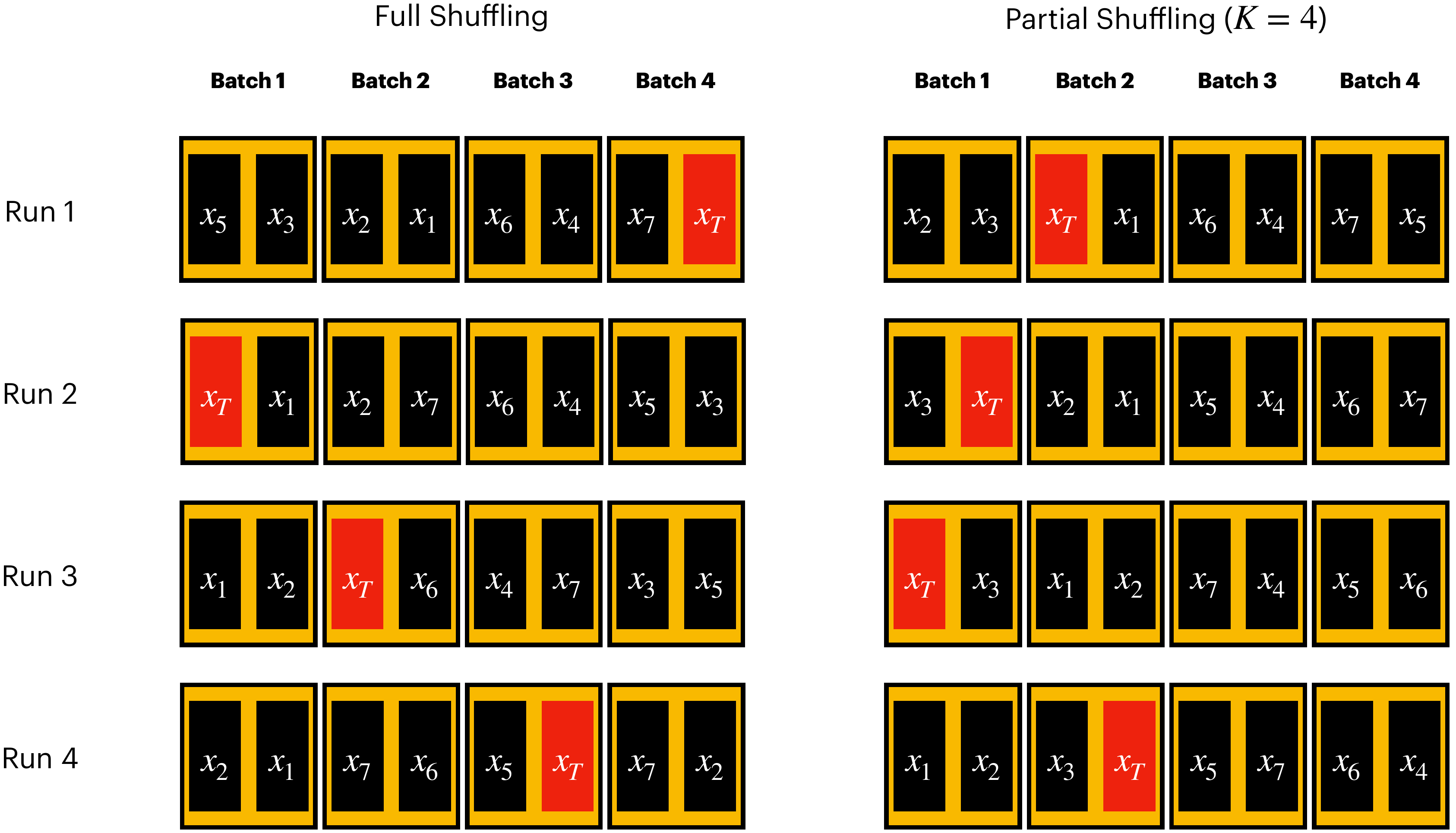}
  \caption{Comparison between a small dataset with eight records ($x_1$,...,$x_7$,$x_T$) being fully shuffled vs.~partially shuffled with a buffer of $K{=}4$ and batch size $B{=}2$. The target record $x_T$ is highlighted in red.} %
  \label{fig:viz_partial_shuffle}
\end{figure}

We first consider the case where datasets are only shuffled within a small buffer of $K$ samples, as reported in~\cite{ponomareva2023dp}, 
and report the procedure used to debug this variation in Algorithm~\ref{alg:audit_partial_shuffle}.
The dataset is shuffled in buffers of size $K$, i.e., the first $K$ samples are shuffled and batched, followed by the next $K$ samples, etc.
Figure~\ref{fig:viz_partial_shuffle} provides a visualization of a small one-dimensional dataset that is partially shuffled.

\descr{Methodology.}
We assume the \PI adversary, where the output remains (mostly) the same at $(-1, ..., +1, ..., -1) + \mathcal{N}(0, \sigma^2\mathbb{I})$ and $(-1, ..., 0, ..., -1) + \mathcal{N}(0, \sigma^2\mathbb{I})$ for $D$ and $D'$, respectively.
However, while in full shuffling, the `+1' and `0' can appear uniformly in any batch, here they can only appear in the first $\lfloor \frac{K}{B} \rfloor$ batches since the first $K$ samples are first shuffled and then batched.

The remaining challenge in identifying this bug is that the adversary does not have access to the batch sampler $\mathcal{B}$ and, by extension, does not know $K$.
We impose this restriction as, even in a real-world auditing setting, it may be cumbersome for an auditor to be given access to the batch sampler separately---even then, the model trainer may not faithfully use the batch sampler.
Therefore, we allow the adversary to ``guess'' multiple buffer sizes and evaluate empirical privacy leakage from only the first $K$ batches.
The adversary then outputs the maximum empirical privacy leakage observed across all buffer sizes guessed.
Theoretically, they could perform a binary search to reduce the number of guesses made; however, we assume they search over $K = \{1, 10, 20, ..., 100\}$ since the number of batches is small.
Also, in theory, the empirical privacy leakage from each guess must be calculated on separate sets of observations for the 95\% CI to be valid, similar to choosing the optimal threshold.
However, to evaluate the maximum empirical privacy leakage achievable, we omit this step and instead report standard deviations over five runs.

We review the procedure used to debug this variation in Algorithm~\ref{alg:audit_partial_shuffle}.
We denote with $\text{DP-SGD}^{PI}_K$ the execution of DP-SGD with partial shuffle over a buffer of size $K$ and with the modifications made for the \PI adversary as reported in Algorithm~\ref{alg:dpsgd_audit}.

\begin{figure}[t]
  \centering
  \includegraphics[width=\mywidth\linewidth]{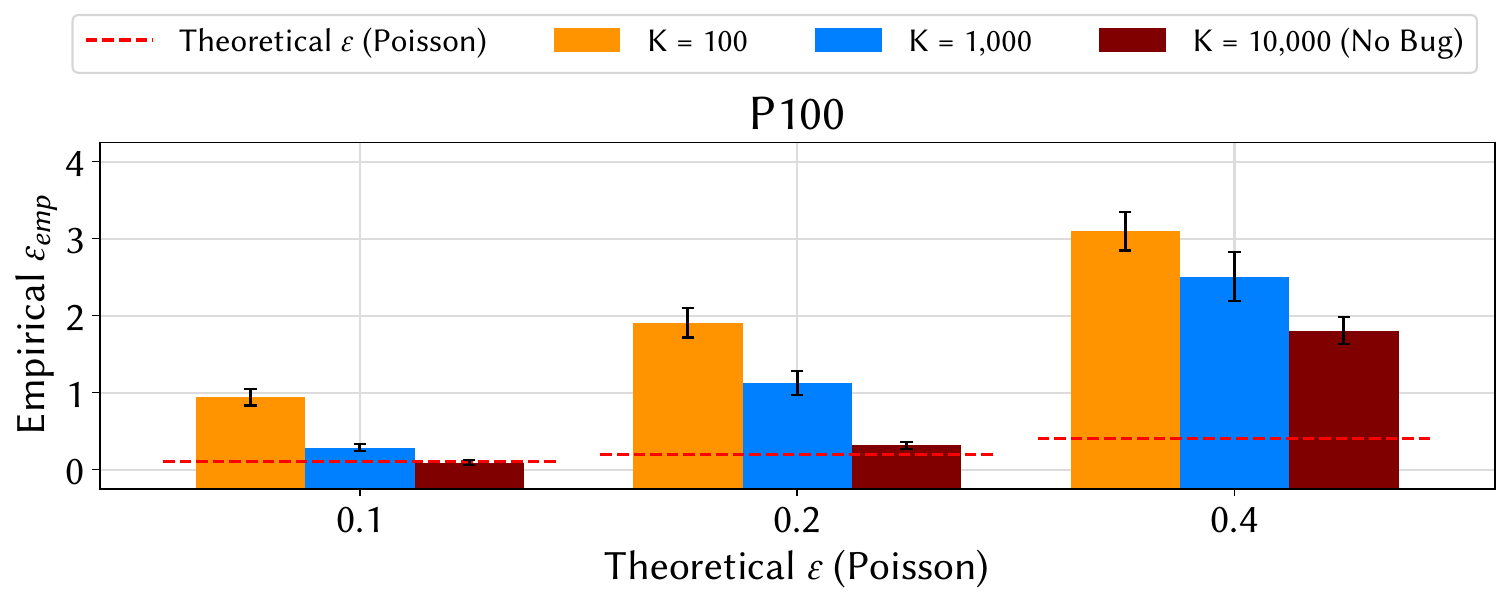}
  \caption{Auditing DP-SGD (Shuffle) when the P100 dataset is only shuffled within a buffer of $K$ samples. When the buffer $K = |D| = 10{,}000$, the dataset is fully shuffled.}
  \label{fig:bug1_exp}
\end{figure}

\descr{Results.}
In Figure~\ref{fig:bug1_exp}, we report the empirical privacy leakage estimates for various buffer sizes $K$.
We only audit the MLP model trained on P100 due to computational constraints.
We experiment with small $\varepsilon = 0.1, 0.2, 0.4$ as the violations are much more significant in this regime.
Specifically, at $\varepsilon = 0.1, 0.2, 0.4$, the empirical privacy leakages for full dataset shuffling $K = 10{,}000$ and partial shuffling at $K = 1{,}000$ are $\empeps = 0.09, 0.31, 1.80$ and $\empeps = 0.29, 1.12, 2.51$, respectively. 
This shows that even partial shuffling can result in substantially larger empirical privacy leakages.

However, as $\varepsilon$ increases, the gap between full shuffling and no shuffling ($K = 100$) reduces.
For instance, at $\varepsilon = 0.1, 0.2, 0.4$, the $\empeps$ values from no shuffling are 10.4$\times$, 6.16$\times$, and 1.72$\times$ that of the $\empeps$ values from full shuffling, respectively.
This indicates that for larger values of $\varepsilon$, the impact of not shuffling or partial shuffling steadily decreases, which in turn suggests that shuffling may not always result in strong privacy amplification.

\subsection{Batch-then-Shuffle}\label{sec:bts}

Finally, we debug the variation where a dataset is first batched and then shuffled.
Although this is not necessarily a known bug in the context of training private models, we find a substantial presence of this sequence of events in public repositories related to non-private ones.
Specifically, we use the GitHub Code Search tool to search for occurrences of \texttt{dataset.batch(.*)\allowbreak.shuffle(.*)}, which indicates that the dataset is first batched before it is shuffled, and compare with occurrences of \texttt{dataset.\allowbreak shuffle(.*).batch(.*)}.

We find approximately 10,800 files using the ``correct'' shuffle-then-batch implementation and 290 files (2.6\%) using the batch-then-shuffle approach.
As a result, we set out to pre-emptively explore techniques to catch these bugs.

\begin{figure}[t]
  \centering
  \includegraphics[width=\linewidth]{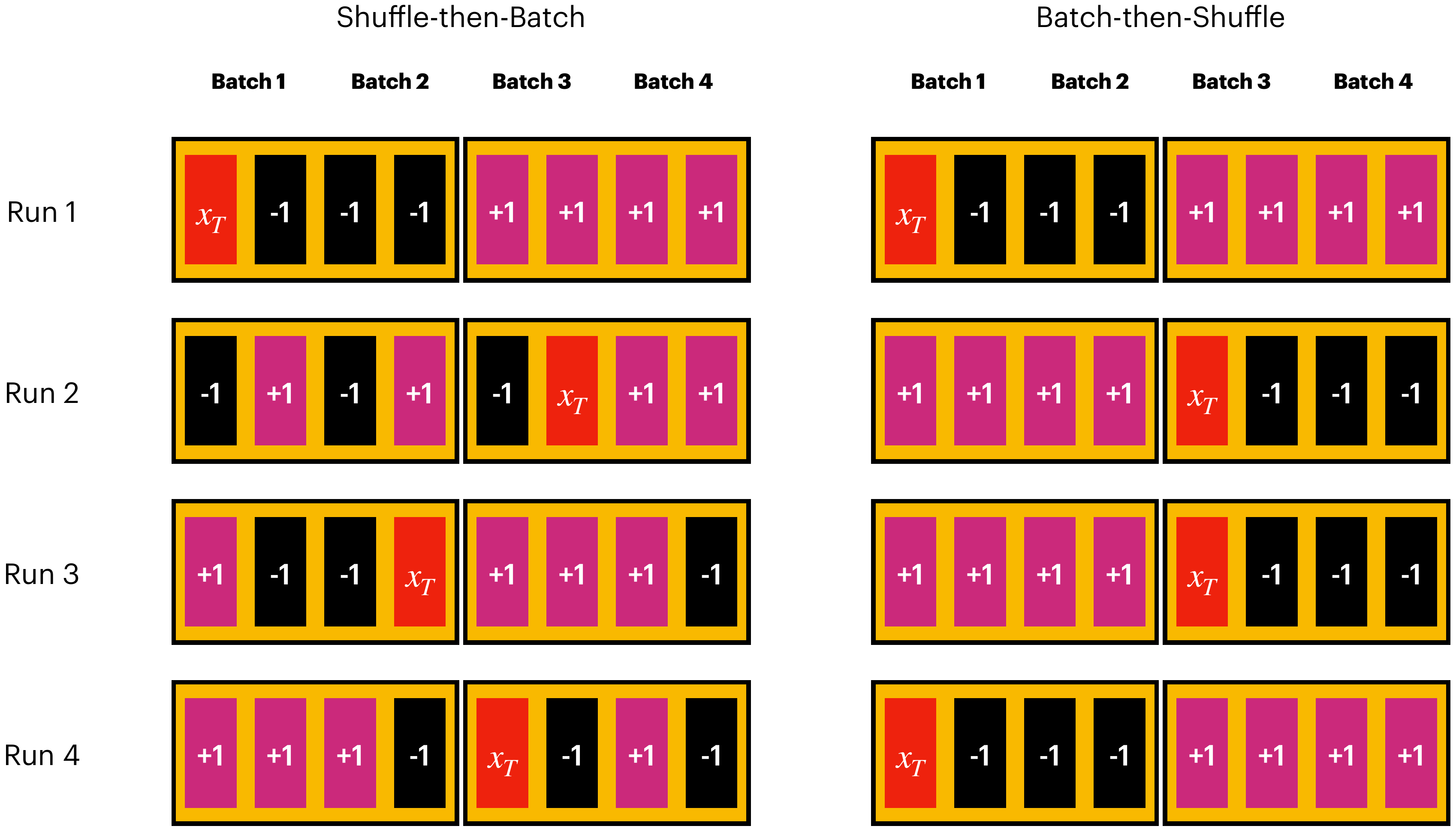}
  \caption{Comparison between a small dataset with 8 records being shuffled then batched (``correct'') vs. batched then shuffled (``wrong'') with batch size $B = 4$. Target record $x_T$ is highlighted in red.}
  \label{fig:viz_batch_then_shuffle}
\end{figure}

Under the \Nat and \PI threat models, batch-then-shuffle is equivalent to shuffle-then-batch.
This is because we expect the dot product of the privatized gradient and gradient of almost all other samples, which cannot be modified by the adversary, to be near 0 under both threat models.
Thus, this bug can only be detected under the \WC threat model.
We then consider the \WC threat model, but we let the adversary insert the gradients of the first $B$ samples as the canary gradient $\hat{g}$ and the remaining samples as the negative canary gradient $-\hat{g}$.
By doing so, the outputs from the DP-SGD algorithm are expected to be $(B, -B, ..., -B) + \mathcal{N}(0, \sigma^2\mathbb{I})$ and $(B - 1, -B, ..., -B) + \mathcal{N}(0, \sigma^2\mathbb{I})$ for $D$ and $D'$, respectively.
Note that `$B$' and `$B - 1$' can appear uniformly in any batch.
Please refer to Figure~\ref{fig:viz_batch_then_shuffle} for a visualization of a small one-dimensional dataset that is batched first and then shuffled.

\begin{figure}[t]
  \centering
  \includegraphics[width=\mywidth\linewidth]{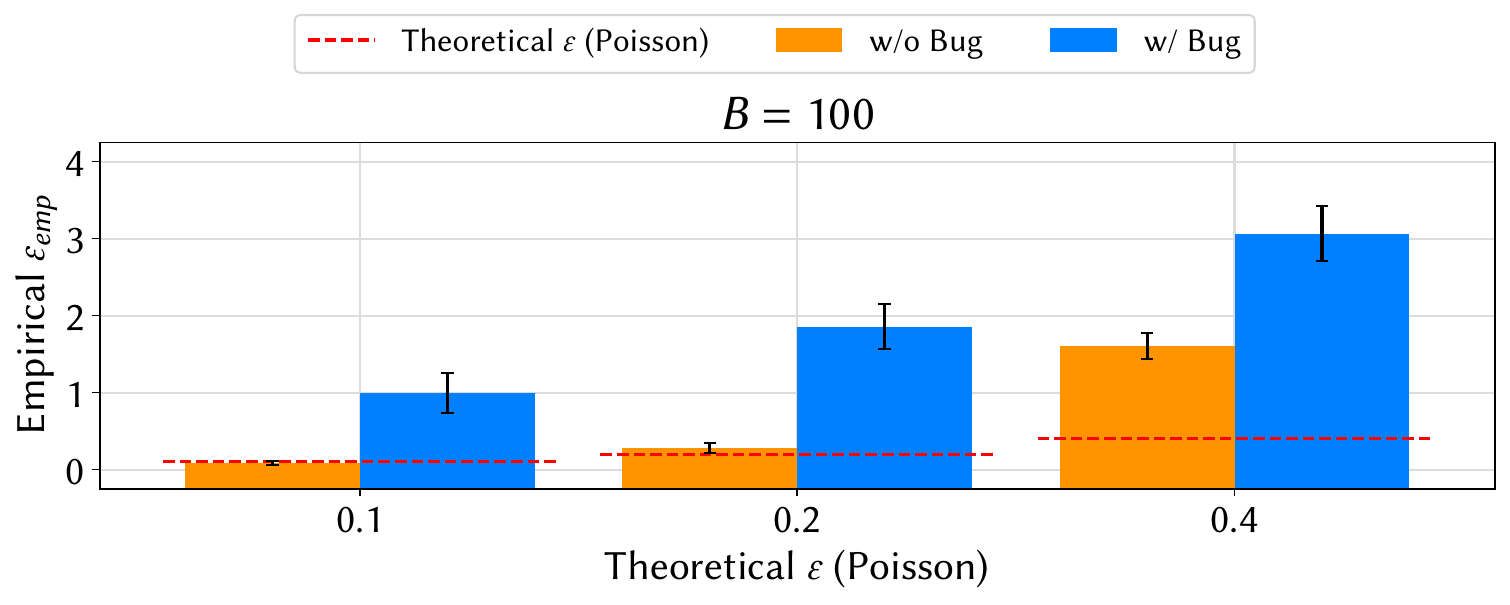}
  \caption{Auditing DP-SGD (Shuffle) when the P100 dataset is batched before shuffling is performed.}
  \label{fig:bug2_exp}
\end{figure}

We adjust the likelihood ratio used under the \WC threat model accordingly and report the audit results with/without the bug in Figure~\ref{fig:bug2_exp}.
Our audit can indeed easily detect this bug: at $\varepsilon = 0.1, 0.2, 0.4$, $\empeps = 1.00, 1.86, 3.06$, respectively, with the bug but only $\empeps = 0.09, 0.28, 1.61$, respectively, without the bug.

\subsection{Takeaways}
Our experiments show that our auditing procedure is highly extensible as it can be used to audit common variations of the shuffling procedure, more precisely, ``partial shuffling'' and ``batch-then-shuffle.'' %
These variations remain easily detectable by our auditing method, which estimates $\empeps = 0.29, 1.12, 2.51$ and $\empeps = 1.00, 1.86, 3.06$ for the ``partial shuffling'' and ``batch-then-shuffle'' procedure at theoretical $\varepsilon = 0.1, 0.2, 0.4$, respectively.

\section{Related Work}\label{sec:related}
To our knowledge, we are the first to audit DP-SGD with shuffling. %
In the following, we review relevant prior work on shuffling and auditing DP-SGD.

\descr{DP-SGD Audits.} Jayaraman et al.~\cite{jayaraman2019evaluating} present the first DP-SGD audit using black-box inference attacks on the trained model, but only achieve loose estimates of the privacy leakage.
Jagielski et al.~\cite{jagielski2020auditing} use input canaries to improve tightness and use Clopper-Pearson intervals to calculate confidence intervals for the empirical estimates. %
Nasr et al.~\cite{nasr2021adversary} are the first to achieve tight DP-SGD audits using a large number of training runs, a pathological dataset, and an active white-box adversary that can insert gradient canaries at each step.
To reduce the number of training runs, Zanella-B{\'e}guelin et al.~\cite{zanella2023bayesian} use credible intervals instead of Clopper-Pearson.
Although~\cite{nasr2023tight} later questions the validity of the credible intervals, it shows that tight empirical estimates are possible even with few training runs and natural (not adversarially crafted) datasets by auditing with $f$-DP.
Recent work on DP-SGD audits has also focused on federated learning settings~\cite{maddock2023canife,galen2024oneshot} and further reducing the number of training runs~\cite{pillutla2024unleashing} to a single one~\cite{galen2024oneshot,steinke2024privacy,mahloujifar2024auditing} and auditing under weak threat models~\cite{annamalai2024nearly,cebere2024tighter}.

Overall, prior work auditing DP-SGD implementations has focused on Poisson subsampling.
Previous audits also typically threshold the losses~\cite{annamalai2024nearly,jagielski2020auditing,mahloujifar2024auditing,nasr2023tight,nasr2021adversary,pillutla2024unleashing,steinke2024privacy} or gradients~\cite{mahloujifar2024auditing,nasr2023tight,nasr2021adversary,nasr2021adversary,pillutla2024unleashing} directly.
By contrast, we audit the shuffling setting and use likelihood ratio functions to audit DP-SGD.

\descr{Shuffling and DP-SGD.} To the best of our knowledge, there is limited prior work analyzing the privacy of DP-SGD with shuffling.
Chua et al.~\cite{chua2024private} analyze a simplified version of DP-SGD, i.e., the Adaptive Batch Linear Query (ABLQ) mechanism.
Although they do not present a theoretical upper bound for the ABLQ mechanism with shuffling, they consider pathological settings where ABLQ could leak more privacy when using shuffling instead of Poisson subsampling.
While they initially only consider a single epoch~\cite{chua2024private}, in follow-up work~\cite{chua2024scalable}, they extend their lower bound to multiple epochs.
However, neither work directly analyzes DP-SGD or evaluates the gap in real-world parameter settings. %

Our work takes a different approach, empirically estimating the privacy leakage from DP-SGD using DP auditing.
By doing so, we derive empirical privacy leakage estimates for real-world models trained using DP-SGD (Shuffle).
We also focus on the impact of real-world considerations like threat models and variations on the shuffling procedure.
Overall, we believe that our work is orthogonal to~\cite{chua2024private} as our focus is on \emph{empirical} privacy estimation, whereas theirs is on \emph{theoretical} privacy analysis.

\descr{Shuffling.}
Shuffling is also used in local DP, where users randomize their inputs before sharing them for data processing.
Bittau et al.~\cite{bittau2017prochlo} introduce the Encode-Shuffle-Analyze (ESA) framework, where users' randomized samples are shuffled before being processed.
Cheu et al.~\cite{cheu2019distributed} and Erlingsson et al.~\cite{erlingsson2019amplification} prove that shuffling users' data improves privacy guarantees in the local DP setting.
In follow-up work, Erlingsson et al.~\cite{erlingsson2020encode} apply their privacy amplification results to the ESA framework, while Balle et al.~\cite{balle2019privacy} improve on and generalize results from~\cite{erlingsson2019amplification}.
Finally, Feldman et al.~\cite{feldman2022hiding,feldman2023stronger} and Wang et al.~\cite{wang2023unified} present a nearly optimal analysis of shuffling in the local DP setting.
Overall, the privacy analysis for \approxdp mechanisms typically used in central DP is far less understood than that of shuffling for local DP mechanisms, which typically guarantee pure DP (i.e., $\delta = 0$).

\section{Discussion and Conclusion}\label{sec:conclusion}

This paper presented the first audit of %
DP-SGD implementations that shuffle the training data and deterministically iterate over fixed-size batches, which we denoted as DP-SGD (Shuffle).
Arguably, this addressed an important research problem as using shuffling to sample batches in DP-SGD has become common~\cite{ponomareva2023dp,de2022unlocking,LTLH22}, mostly due to better computational efficiency, even though theoretical guarantees are reported as if Poisson subsampling was used, which we referred to as DP-SGD (Poisson).
This makes it crucial to investigate the gap between the empirical privacy leakage from the former and the theoretical guarantees of the latter.

We introduced new auditing techniques and audited DP-SGD (Shuffle) with different parameter settings and threat models.
More precisely, we experimented with 101 settings from SOTA models using shuffling~\cite{de2022unlocking,LTLH22}, finding gaps in two-thirds of them---in some cases, the empirical privacy leakage from DP-SGD (Shuffle) was up to 4 times higher than the theoretical guarantee from DP-SGD (Poisson).
We also showed how these discrepancies %
can depend on the batch size and the strength of the adversary, with smaller batch sizes and stronger threat models yielding larger gaps.
Finally, we used our techniques to detect the presence of bugs in the implementations of the shuffling procedure.
We investigated two common bugs~\cite{ponomareva2023dp} found in public code repositories of non-private model training, showing that our auditing framework effectively identifies them and detects even higher empirical privacy leakages in their presence. %

\descr{Implications and Recommendations for Practitioners.} The measurable differences between the empirical and the theoretical guarantees %
shows that prior work~\cite{de2022unlocking,LTLH22} substantially overestimated the privacy guarantees of some SOTA private models.
Consequently, our work calls into question the validity of reporting theoretical guarantees using DP-SGD (Poisson) while implementing algorithms that use DP-SGD (Shuffle), reiterating the importance of formally analyzing the privacy guarantees provided by shuffling the dataset.
Concretely, we recommend practitioners avoid shuffling until tighter guarantees are proven. 
Recently proposed alternative sampling schemes like Balls-in-Bins~\cite{choquette-choo2025nearexact,chua2025balls} could also be used, as they could balance shuffling's computational efficiency with provably tight guarantees.

\descr{Adversarial Models.} 
Our audits use adversaries in active white-box models; %
specifically, we assume they can insert a target sample of their choice and view the final trained model, and can insert (varying sizes of) canary gradients at each step and observe all intermediate models.
While we introduce an adversarial model unique to shuffling, i.e., \PI, this serves as a hypothetical middle-ground between the \WC adversary and a weaker one (\Nat), and may not necessarily reflect an actual adversary.
Nevertheless, as discussed in Section~\ref{sec:threat_models}, this primarily serves as a useful auditing tool, enabling us to identify the impact that the bias from other (non-target) samples has on our novel likelihood-based auditing method.

Overall, although not all the adversaries we consider may always have the corresponding capabilities against models deployed in production, %
recall that provably correct DP guarantees are meant to be robust in \emph{worst-case} settings, making our adversaries valid in the context of auditing.
In fact, similar assumptions were made in prior DP-SGD audits~\cite{nasr2021adversary,nasr2023tight,annamalai2024what}.

\descr{Limitations \& Future Work.} 
One main limitation of our work is the computational cost required to run the audit.
Specifically, since shuffling does not provide any improved \fdp or \gdp guarantees, we %
require training runs in the order of thousands and millions.
Recent work presented audits of DP-SGD (Poisson) with just one run~\cite{galen2024oneshot,steinke2024privacy,pillutla2024unleashing,mahloujifar2024auditing}; however, these techniques may not directly apply to DP-SGD (Shuffle) and underestimate privacy leakage even under powerful threat models.
Thus, it is not yet clear how to adapt them to our setting.
In follow-up work, we plan to investigate how to minimize the number of training runs required to audit DP-SGD (Shuffle) accurately, which would also enable us to audit larger deep learning models.

In addition, our audits are weaker for larger batch sizes, which we believe is mainly due to the bias introduced by the ``other'' samples.
We plan to reduce this bias term using debiasing techniques (e.g.,~\cite{tong2025how}) to make our audits effective even for large batch sizes.

Finally, due to computational constraints, we only audited one epoch of DP-SGD (Shuffle) and therefore did not measure model utility.
In the future, we plan to audit state-of-the-art models built using DP-SGD (Shuffle) directly, which will also allow us to measure drops in utility with different adversaries.
As mentioned, we expect utility to plummet with \WC but remain stable with \PI and \Nat adversaries.

\descr{Acknowledgements.}
This work has been supported by the National Science Scholarship (PhD) from the Agency for Science Technology and Research, Singapore.

{\small
\bibliographystyle{abbrv}

\begin{thebibliography}{10}

\bibitem{abadi2016deep}
M.~Abadi, A.~Chu, I.~Goodfellow, H.~B. McMahan, I.~Mironov, K.~Talwar, and
  L.~Zhang.
\newblock {Deep Learning with Differential Privacy}.
\newblock In {\em CCS}, 2016.

\bibitem{galen2024oneshot}
G.~Andrew, P.~Kairouz, S.~Oh, A.~Oprea, H.~B. McMahan, and V.~Suriyakumar.
\newblock {One-shot Empirical Privacy Estimation for Federated Learning}.
\newblock In {\em ICLR}, 2024.

\bibitem{annamalai2024nearly}
M.~S. M.~S. Annamalai and E.~De~Cristofaro.
\newblock {Nearly Tight Black-Box Auditing of Differentially Private Machine
  Learning}.
\newblock In {\em NeurIPS}, 2024.

\bibitem{annamalai2024what}
M.~S. M.~S. Annamalai, G.~Ganev, and E.~De~Cristofaro.
\newblock {``What do you want from theory alone?'' Experimenting with Tight
  Auditing of Differentially Private Synthetic Data Generation}.
\newblock In {\em USENIX Security}, 2024.

\bibitem{balle2018privacy}
B.~Balle, G.~Barthe, and M.~Gaboardi.
\newblock {Privacy Amplification by Subsampling: Tight Analyses via Couplings
  and Divergences}.
\newblock In {\em NeurIPS}, 2018.

\bibitem{balle2019privacy}
B.~Balle, J.~Bell, A.~Gasc{\'o}n, and K.~Nissim.
\newblock {The Privacy Blanket of the Shuffle Model}.
\newblock In {\em CRYPTO}, 2019.

\bibitem{jax-privacy2022github}
B.~Balle, L.~Berrada, S.~De, S.~Ghalebikesabi, J.~Hayes, A.~Pappu, S.~L. Smith,
  and R.~Stanforth.
\newblock {JAX-Privacy: Algorithms for Privacy-Preserving Machine Learning in
  JAX}.
\newblock \url{https://github.com/google-deepmind/jax_privacy}, 2022.

\bibitem{bittau2017prochlo}
A.~Bittau, {\'U}.~Erlingsson, P.~Maniatis, I.~Mironov, A.~Raghunathan, D.~Lie,
  M.~Rudominer, U.~Kode, J.~Tinnes, and B.~Seefeld.
\newblock {Prochlo: Strong Privacy for Analytics in the Crowd}.
\newblock In {\em SOPS}, 2017.

\bibitem{carlini2022membership}
N.~Carlini, S.~Chien, M.~Nasr, S.~Song, A.~Terzis, and F.~Tram{\`e}r.
\newblock {Membership Inference Attacks From First Principles}.
\newblock In {\em IEEE S\&P}, 2022.

\bibitem{opacusbug}
T.~Cebere.
\newblock {Privacy Leakage at low sample size}.
\newblock \url{https://github.com/pytorch/opacus/issues/571}, 2023.

\bibitem{cebere2024tighter}
T.~Cebere, A.~Bellet, and N.~Papernot.
\newblock {Tighter Privacy Auditing of DP-SGD in the Hidden State Threat
  Model}.
\newblock {\em arXiv:2405.14457}, 2024.

\bibitem{cheu2019distributed}
A.~Cheu, A.~Smith, J.~Ullman, D.~Zeber, and M.~Zhilyaev.
\newblock {Distributed Differential Privacy via Shuffling}.
\newblock In {\em Eurocrypt}, 2019.

\bibitem{choquette-choo2025nearexact}
C.~A. Choquette-Choo, A.~Ganesh, S.~Haque, T.~Steinke, and A.~G. Thakurta.
\newblock {Near-Exact Privacy Amplification for Matrix Mechanisms}.
\newblock In {\em ICLR}, 2025.

\bibitem{choquette2024amplified}
C.~A. Choquette-Choo, A.~Ganesh, R.~McKenna, H.~B. McMahan, J.~Rush,
  A.~Guha~Thakurta, and Z.~Xu.
\newblock {(Amplified) Banded Matrix Factorization: A unified approach to
  private training}.
\newblock In {\em NeurIPS}, 2024.

\bibitem{chua2025balls}
L.~Chua, B.~Ghazi, C.~Harrison, E.~Leeman, P.~Kamath, R.~Kumar, P.~Manurangsi,
  A.~Sinha, and C.~Zhang.
\newblock {Balls-and-bins sampling for DP-SGD}.
\newblock In {\em AISTATS}, 2025.

\bibitem{chua2024private}
L.~Chua, B.~Ghazi, P.~Kamath, R.~Kumar, P.~Manurangsi, A.~Sinha, and C.~Zhang.
\newblock {How Private are DP-SGD Implementations?}
\newblock In {\em ICML}, 2024.

\bibitem{chua2024scalable}
L.~Chua, B.~Ghazi, P.~Kamath, R.~Kumar, P.~Manurangsi, A.~Sinha, and C.~Zhang.
\newblock {Scalable DP-SGD: Shuffling vs. Poisson Subsampling}.
\newblock {\em arXiv:2411.04205}, 2024.

\bibitem{clopper1934use}
C.~J. Clopper and E.~S. Pearson.
\newblock {The use of confidence or fiducial limits illustrated in the case of
  the binomial}.
\newblock {\em Biometrika}, 1934.

\bibitem{de2022unlocking}
S.~De, L.~Berrada, J.~Hayes, S.~L. Smith, and B.~Balle.
\newblock {Unlocking High-Accuracy Differentially Private Image Classification
  through Scale}.
\newblock {\em arXiv:2204.13650}, 2022.

\bibitem{debenedetti2024privacy}
E.~Debenedetti, G.~Severi, N.~Carlini, C.~A. Choquette-Choo, M.~Jagielski,
  M.~Nasr, E.~Wallace, and F.~Tram{\`e}r.
\newblock {Privacy Side Channels in Machine Learning Systems}.
\newblock In {\em USENIX}, 2024.

\bibitem{ding2018detecting}
Z.~Ding, Y.~Wang, G.~Wang, D.~Zhang, and D.~Kifer.
\newblock {Detecting Violations of Differential Privacy}.
\newblock In {\em CCS}, 2018.

\bibitem{dong2019gaussian}
J.~Dong, A.~Roth, and W.~J. Su.
\newblock {Gaussian Differential Privacy}.
\newblock {\em arXiv:1905.02383}, 2019.

\bibitem{dormann2021not}
F.~D{\"o}rmann, O.~Frisk, L.~N. Andersen, and C.~F. Pedersen.
\newblock {Not All Noise is Accounted Equally: How Differentially Private
  Learning Benefits from Large Sampling Rates}.
\newblock In {\em MLSP}, 2021.

\bibitem{dwork2006calibrating}
C.~Dwork, F.~McSherry, K.~Nissim, and A.~Smith.
\newblock {Calibrating Noise to Sensitivity in Private Data Analysis}.
\newblock In {\em TCC}, 2006.

\bibitem{erlingsson2020encode}
{\'U}.~Erlingsson, V.~Feldman, I.~Mironov, A.~Raghunathan, S.~Song, K.~Talwar,
  and A.~Thakurta.
\newblock {Encode, Shuffle, Analyze Privacy Revisited: Formalizations and
  Empirical Evaluation}.
\newblock {\em arXiv:2001.03618}, 2020.

\bibitem{erlingsson2019amplification}
{\'U}.~Erlingsson, V.~Feldman, I.~Mironov, A.~Raghunathan, K.~Talwar, and
  A.~Thakurta.
\newblock {Amplification by Shuffling: From Local to Central Differential
  Privacy via Anonymity}.
\newblock In {\em SODA}, 2019.

\bibitem{feldman2022hiding}
V.~Feldman, A.~McMillan, and K.~Talwar.
\newblock {Hiding Among the Clones: A Simple and Nearly Optimal Analysis of
  Privacy Amplification by Shuffling}.
\newblock In {\em FOCS}, 2022.

\bibitem{feldman2023stronger}
V.~Feldman, A.~McMillan, and K.~Talwar.
\newblock {Stronger Privacy Amplification by Shuffling for Renyi and
  Approximate Differential Privacy}.
\newblock In {\em SODA}, 2023.

\bibitem{feldman2025privacy}
V.~Feldman and M.~Shenfeld.
\newblock {Privacy amplification by random allocation}.
\newblock {\em arXiv:2502.08202}, 2025.

\bibitem{tfprivacy}
Google.
\newblock {TensorFlow Privacy}.
\newblock \url{https://github.com/tensorflow/privacy}, 2019.

\bibitem{jagielski2020auditing}
M.~Jagielski, J.~Ullman, and A.~Oprea.
\newblock {Auditing Differentially Private Machine Learning: How Private is
  Private SGD?}
\newblock In {\em NeurIPS}, 2020.

\bibitem{jayaraman2019evaluating}
B.~Jayaraman and D.~Evans.
\newblock {Evaluating differentially private machine learning in practice}.
\newblock In {\em USENIX Security}, 2019.

\bibitem{prngreuse}
M.~Johnson.
\newblock {Fix prng key reuse in differential privacy example}.
\newblock \url{https://github.com/google/jax/pull/3646}, 2020.

\bibitem{kairouz2021practical}
P.~Kairouz, B.~McMahan, S.~Song, O.~Thakkar, A.~Thakurta, and Z.~Xu.
\newblock {Practical and Private (Deep) Learning Without Sampling or
  Shuffling}.
\newblock In {\em ICML}, 2021.

\bibitem{kairouz2015composition}
P.~Kairouz, S.~Oh, and P.~Viswanath.
\newblock {The Composition Theorem for Differential Privacy}.
\newblock In {\em ICML}, 2015.

\bibitem{koskela2020computing}
A.~Koskela, J.~J{\"a}lk{\"o}, and A.~Honkela.
\newblock {Computing Tight Differential Privacy Guarantees Using FFT}.
\newblock In {\em AISTATS}, 2020.

\bibitem{krizhevsky2009learning}
A.~Krizhevsky.
\newblock {Learning Multiple Layers of Features from Tiny Images}.
\newblock \url{https://www.cs.utoronto.ca/~kriz/learning-features-2009-TR.pdf},
  2009.

\bibitem{lecun1998gradient}
Y.~LeCun, L.~Bottou, Y.~Bengio, and P.~Haffner.
\newblock Gradientbased learning applied to document recognition.
\newblock {\em Proceedings of the IEEE}, 86(11), 1998.

\bibitem{LTLH22}
X.~Li, F.~Tram{\`e}r, P.~Liang, and T.~Hashimoto.
\newblock {Large Language Models Can Be Strong Differentially Private
  Learners}.
\newblock In {\em ICLR}, 2022.

\bibitem{maddock2023canife}
S.~Maddock, A.~Sablayrolles, and P.~Stock.
\newblock {CANIFE: Crafting Canaries for Empirical Privacy Measurement in
  Federated Learning}.
\newblock In {\em ICLR}, 2023.

\bibitem{mahloujifar2024auditing}
S.~Mahloujifar, L.~Melis, and K.~Chaudhuri.
\newblock {Auditing $f$-Differential Privacy in One Run}.
\newblock In {\em ICML}, 2025.

\bibitem{nasr2023tight}
M.~Nasr, J.~Hayes, T.~Steinke, B.~Balle, F.~Tram{\`e}r, M.~Jagielski,
  N.~Carlini, and A.~Terzis.
\newblock {Tight Auditing of Differentially Private Machine Learning}.
\newblock In {\em USENIX Security}, 2023.

\bibitem{nasr2021adversary}
M.~Nasr, S.~Songi, A.~Thakurta, N.~Papernot, and N.~Carlini.
\newblock {Adversary Instantiation: Lower Bounds for Differentially Private
  Machine Learning}.
\newblock In {\em IEEE S\&P}, 2021.

\bibitem{neyman1933ix}
J.~Neyman and E.~S. Pearson.
\newblock {IX. On the problem of the most efficient tests of statistical
  hypotheses}.
\newblock {\em Philosophical Transactions of the Royal Society of London.
  Series A, Containing Papers of a Mathematical or Physical Character},
  231(694-706):289--337, 1933.

\bibitem{xla}
OpenXLA.
\newblock {XLA (Accelerated Linear Algebra)}.
\newblock \url{https://github.com/openxla/xla}, 2024.

\bibitem{pillutla2024unleashing}
K.~Pillutla, G.~Andrew, P.~Kairouz, H.~B. McMahan, A.~Oprea, and S.~Oh.
\newblock {Unleashing the Power of Randomization in Auditing Differentially
  Private ML}.
\newblock In {\em NeurIPS}, 2024.

\bibitem{ponomareva2023dp}
N.~Ponomareva, S.~Vassilvitskii, Z.~Xu, B.~McMahan, A.~Kurakin, and C.~Zhang.
\newblock {How to DP-fy ML: A Practical Tutorial to Machine Learning with
  Differential Privacy}.
\newblock In {\em KDD}, 2023.

\bibitem{shamsabadi2024confidential}
A.~S. Shamsabadi, G.~Tan, T.~I. Cebere, A.~Bellet, H.~Haddadi, N.~Papernot,
  X.~Wang, and A.~Weller.
\newblock {Confidential-DPproof: Confidential Proof of Differentially Private
  Training}.
\newblock In {\em ICLR}, 2024.

\bibitem{shokri2017membership}
R.~Shokri, M.~Stronati, C.~Song, and V.~Shmatikov.
\newblock {Membership Inference Attacks against Machine Learning Models}.
\newblock In {\em IEEE S\&P}, 2017.

\bibitem{steinke2024privacy}
T.~Steinke, M.~Nasr, and M.~Jagielski.
\newblock {Privacy Auditing with One (1) Training Run}.
\newblock In {\em NeurIPS}, 2024.

\bibitem{tong2025how}
Y.~Tong, J.~Ye, S.~Zarifzadeh, and R.~Shokri.
\newblock {How much of my dataset did you use? Quantitative Data Usage
  Inference in Machine Learning}.
\newblock In {\em ICLR}, 2025.

\bibitem{tramer2021differentially}
F.~Tram{\`e}r and D.~Boneh.
\newblock {Differentially Private Learning Needs Better Features (or Much More
  Data)}.
\newblock In {\em ICLR}, 2021.

\bibitem{tramer2022debugging}
F.~Tram{\`e}r, A.~Terzis, T.~Steinke, S.~Song, M.~Jagielski, and N.~Carlini.
\newblock {Debugging Differential Privacy: A Case Study for Privacy Auditing}.
\newblock {\em arXiv:2202.12219}, 2022.

\bibitem{wang2023unified}
C.~Wang, B.~Su, J.~Ye, R.~Shokri, and W.~Su.
\newblock {Unified Enhancement of Privacy Bounds for Mixture Mechanisms via
  $f$-Differential Privacy}.
\newblock In {\em NeurIPS}, 2023.

\bibitem{xiang2025privacy}
Z.~Xiang, T.~Wang, and D.~Wang.
\newblock {Privacy Audit as Bits Transmission:(Im) possibilities for Audit by
  One Run}.
\newblock In {\em USENIX Security}, 2025.

\bibitem{xiao2017fashion}
H.~Xiao, K.~Rasul, and R.~Vollgraf.
\newblock {Fashion-MNIST: a Novel Image Dataset for Benchmarking Machine
  Learning Algorithms}.
\newblock {\em arXiv:1708.07747}, 2017.

\bibitem{ye2022enhanced}
J.~Ye, A.~Maddi, S.~K. Murakonda, V.~Bindschaedler, and R.~Shokri.
\newblock {Enhanced Membership Inference Attacks against Machine Learning
  Models}.
\newblock In {\em CCS}, 2022.

\bibitem{yousefpour2021opacus}
A.~Yousefpour, I.~Shilov, A.~Sablayrolles, D.~Testuggine, K.~Prasad, M.~Malek,
  J.~Nguyen, S.~Ghosh, A.~Bharadwaj, J.~Zhao, et~al.
\newblock {Opacus: User-friendly Differential Privacy Library in PyTorch}.
\newblock {\em arXiv:2109.12298}, 2021.

\bibitem{zanella2023bayesian}
S.~Zanella-B{\'e}guelin, L.~Wutschitz, S.~Tople, A.~Salem, V.~R{\"u}hle,
  A.~Paverd, M.~Naseri, B.~K{\"o}pf, and D.~Jones.
\newblock {Bayesian Estimation of Differential Privacy}.
\newblock In {\em ICML}, 2023.

\bibitem{zhu2022optimal}
Y.~Zhu, J.~Dong, and Y.-X. Wang.
\newblock {Optimal Accounting of Differential Privacy via Characteristic
  Function}.
\newblock In {\em AISTATS}, 2022.

\end{thebibliography}

}

\appendix

\section{Differentially Private Stochastic Gradient Descent (DP-SGD)}\label{app:dpsgd}
In Algorithm~\ref{alg:dpsgd}, we review DP-SGD~\cite{abadi2016deep}'s pseudo-code.
Then, in Algorithm~\ref{alg:simple_mech}, we report that of \SMFull (\SM), a heavily simplified version of DP-SGD adapted from~\cite{chua2024private} to develop principled tight auditing techniques for shuffling under an idealized setting.

\begin{algorithm}[b]
    \small
    \caption{Differentially Private Stochastic Gradient Descent (DP-SGD)~\cite{abadi2016deep}%
    }\label{alg:dpsgd}
    \begin{algorithmic}[1]
    \Require Dataset, $D$. Epochs, $E$. Batch Size, $B$. Learning rate, $\eta$. Batch sampler, $\mathcal{B}$. Loss function, $\ell$. Initial model parameters, $\theta_0$. Noise multiplier, $\sigma$. Clipping norm, $C$.
    \State $T \leftarrow |D| / B$
    \For{$i \in [E]$}
      \State $\theta^i_1 \leftarrow \theta^{i - 1}_T$
      \State Sample batches $B_1, ..., B_T \leftarrow \mathcal{B}(D, B)$
      \For{$t \in [T]$}
        \For{$(x_j, y_j) \in B_t$}
            \State $g_j \leftarrow \nabla \ell((x_j, y_j); \theta^i_t)$
            \State $\bar{g}_j \leftarrow g_j / \max(1, \frac{||g_j||_2}{C})$
        \EndFor
        \State $\tilde{g} \leftarrow \frac{1}{B} \left(\sum_j \bar{g}_j + \mathcal{N}(0, C^2\sigma^2\mathbb{I})\right)$
        \State $\theta^i_{t + 1} \leftarrow \theta^i_t - \eta\tilde{g}$
      \EndFor
    \EndFor
    \State \Return $\theta^E_T$
    \color{black}
    \end{algorithmic}
\end{algorithm}

\begin{algorithm}[t]
    \small
    \caption{\SMFull (\SM)}\label{alg:simple_mech}
    \begin{algorithmic}[1]
    \Require Dataset, $D = (x_1, ..., x_N) \in [-1, +1]^N$. Batch Size, $B$. Number of epochs, $E$. Batch sampler, $\mathcal{B}$. Noise multiplier, $\sigma$.
    \State $T \leftarrow |D| / B$
    \For{$i \in [E]$}
      \State Sample batches $B_1, ..., B_T \leftarrow \mathcal{B}(D, B)$
      \For{$t \in [T]$}
        \State $\tilde{g}^i_t \leftarrow \sum_{x_i \in B_t}  x_i + \mathcal{N}(0, \sigma^2)$
      \EndFor
    \EndFor
    \State \Return $\begin{bmatrix} 
      \tilde{g}^1_1 & \dots  & \tilde{g}^1_T\\
      \vdots & \ddots & \vdots\\
      \tilde{g}^E_1 & \dots  & \tilde{g}^E_T
      \end{bmatrix}$
    \color{black}
    \end{algorithmic}
\end{algorithm}

\begin{table}[t]
	\small
  \begin{minipage}[t]{\linewidth}
	\centering
  \begin{tabular}[t]{ll}
    \toprule
    \textbf{Layer} & \textbf{Parameters}\\
\midrule
    Convolution & 32 filters of 3x3, stride 1, padding 1 \\
    Convolution & 32 filters of 3x3, stride 1, padding 1 \\
    Max-Pooling & 2x2, stride 2, padding 0 \\
    Convolution & 64 filters of 3x3, stride 1, padding 1 \\
    Convolution & 64 filters of 3x3, stride 1, padding 1 \\
    Max-Pooling & 2x2, stride 2, padding 0 \\
    Convolution & 128 filters of 3x3, stride 1, padding 1 \\
    Convolution & 128 filters of 3x3, stride 1, padding 1 \\
    Max-Pooling & 2x2, stride 2, padding 0 \\
    Fully connected & 128 units\\
    Fully connected & 2 units\\
    \bottomrule
  \end{tabular}	
  \caption{Shallow CNN model for CIFAR-10 with Tanh activations.}
  \label{tab:cnn_cifar10}
	\small
	\centering
  \begin{tabular}[t]{ll}
\toprule
    \textbf{Layer} & \textbf{Parameters}\\
    \midrule
    Convolution & 6 filters of 5x5, stride 1, padding 2\\
    Avg-Pooling & 2x2 \\
    Convolution & 16 filters of 5x5, stride 1 \\
    Avg-Pooling & 2x2 \\
    Fully connected & 120 units\\
    Fully connected & 84 units\\
    Fully connected & 2 units\\
    \bottomrule
  \end{tabular}	
\sptable
  \caption{LeNet-5 model for FMNIST with Tanh activations.}
  \label{tab:cnn_fmnist}
    \end{minipage}
\end{table}

\section{Worst-Case vs Pathological Dataset}
\label{app:worst_case}

We now briefly explain how our Worst-Case adversary can be modeled with the Pathological Dataset adversary used in prior work~\cite{nasr2021adversary}.
The Pathological Dataset adversary crafts the malicious dataset by making sure the dataset (not including the target sample) is labeled perfectly by the initial model.
This ensures that the gradients of all other samples are zero, except for the target sample.
Furthermore, to ensure that subsequent model updates do not disrupt the gradients, the Pathological Dataset also sets the learning rate to 0.

On the other hand, in our work, the Worst-Case adversary crafts malicious gradients for all samples that are equal in magnitude but opposite in direction to the target sample.
This can be modeled by the Pathological Dataset adversary by first calculating the gradient of a given target sample with respect to the initial model parameters.
Then, the other samples can be crafted by optimizing an input sample with respect to the target sample's gradient in the opposite direction and repeating them $N$ times.
Finally, the learning rate can be set to 0 to ensure that future model updates do not disrupt the gradients.
Therefore, we can ensure that the gradient of the target sample is always equal in magnitude and opposite in direction to the target sample throughout the training process.

\section{Model Architectures}
\label{app:model_arch}

In our experiments, we audit several (shallow) Convolutional Neural Networks (CNNs) and an MLP model corresponding to different datasets. 
More precisely, for the CIFAR-10 dataset, we use the CNN from D{\"o}rmann et al.~\cite{dormann2021not}, who make minor modifications to the CNNs previously used by Tram{\`e}r and Boneh~\cite{tramer2021differentially}.
For the FMNIST dataset, we use a small LeNet-5 model~\cite{lecun1998gradient}.
Finally, for the P100 dataset, we use an MLP model with 32 hidden neurons and ReLU activations.

Exact model architectures for CIFAR-10 and FMNIST are reported in Tables~\ref{tab:cnn_cifar10} and~\ref{tab:cnn_fmnist}, respectively.

\end{document}